\documentclass[final,5p,times,twocolumn,sort&compress]{elsarticle}

\usepackage{graphicx}
\usepackage{amssymb}
\usepackage{amsmath}
\usepackage{empheq}
\usepackage{comment}
\usepackage{subcaption}
\usepackage{balance}
\usepackage{placeins}
\usepackage{enumerate}
\usepackage[inline,shortlabels]{enumitem}
\usepackage[usenames,dvipsnames]{color}
\usepackage[colorlinks=true,linkcolor=Maroon,citecolor=Maroon,urlcolor=Maroon]{hyperref} 
\usepackage{url}

\usepackage[usenames,dvipsnames]{color} 
\usepackage{floatpag} 					
\usepackage{footmisc} 					
\usepackage{booktabs}                   
\usepackage{balance}                    
\usepackage{soul}                       

\newtheorem{definition}{Definition}
\newtheorem{proposition}{Proposition}

\newcommand{\defref}[1]{Def.\ref{#1}}
\newcommand{\propref}[1]{Prop.\ref{#1}}

\newcommand{\A}{\mathbf{A}} 			     		
\newcommand{\PP}{\mathbf{P}} 		     		
\newcommand{\U}{\mathbf{U}} 			     		
\newcommand{\V}{\mathbf{V}} 			     
\newcommand{\x}{\mathbf{x}} 			     		
\newcommand{\f}{\mathbf{f}}
\newcommand{\p}{\mathbf{p}}
\newcommand{\J}{\mathbf{J}}
\newcommand{\I}{\mathbf{I}}
\newcommand{\K}{\mathbf{K}}

\newcommand{\hr}{\hat{\mathbf{r}}}                            
\newcommand{\hn}{\hat{\mathbf{n}}}                            
\newcommand{\hu}{\hat{\mathbf{u}}}
\newcommand{\hv}{\hat{\mathbf{v}}}

\newcommand{\mw}[1]{\langle #1 \rangle}     		 

\newcommand{\et}{\boldsymbol{\eta}}

\newcommand{\Lam}{\boldsymbol{\Lambda}}      
\newcommand{\Sig}{\boldsymbol{\Sigma}}       
\newcommand{\Gam}{\boldsymbol{\Gamma}}       

\DeclareMathOperator*{\argmax}{arg\,max} 		

\makeatletter
\def\ps@pprintTitle{%
 \let\@oddhead\@empty 
 \let\@evenhead\@empty
 \def\@oddfoot{}%
 \let\@evenfoot\@oddfoot}
 \makeatother

\begin{document}
\begin{frontmatter}

\title{
Illusions of Criticality: Crises Without Tipping Points
}

\author[add1]{Virgile Troude\fnref{equal}}
\author[add1]{Sandro Lera\fnref{corr}}
\author[add1]{Ke Wu\fnref{corr}}
\author[add1]{Didier Sornette\fnref{equal}}

\fntext[equal]{These authors contributed equally to this work.}
\fntext[corr]{Corresponding authors. E-mails: $\{$leras,wuk$\}$@sustech.edu.cn}
\address[add1]{\scriptsize Institute of Risks Analysis, Prediction and Management\\
Southern University of Science and Technology, Shenzhen, China}

\begin{abstract}
Abrupt shifts in ecosystems, brains, markets, and climate are often diagnosed as signs of approaching a tipping point, i.e. a critical bifurcation where stability is lost. 
Here we reveal a broader and more deceptive mechanism: pseudo-bifurcations. 
In stochastic non-normal systems, asymmetric interactions produce transient episodes of apparent instability despite long-term stability. 
We show analytically, numerically, and with empirical evidence from brain dynamics during epileptic seizures that pseudo-bifurcations reproduce the full set of early-warning signals usually taken as proof of proximity to tipping points, including critical slowing down, increased variance, and dimensional collapse. 
Crucially, these false alarms can occur well before any true bifurcation, systematically biasing crisis diagnosis. 
This discovery reframes how abrupt transitions are interpreted across disciplines: what has long been attributed to ``criticality'' may instead reflect the hidden geometry of non-normal dynamics. 
By uncovering this illusion of criticality, we call for a fundamental reassessment of how crises are identified, predicted, and managed in natural, social, and technological systems.
\end{abstract}


\end{frontmatter}

\vskip 1cm

\section*{Introduction}

Many of the dynamics driving complex systems in nature and society are marked by abrupt variations that result in transitions to fundamentally different behaviors \cite{Sornettebook2004, Castellano2009, Sornette2017, Jusup2022}.
In recent decades, the dominant framework for understanding these transitions has centered around the concept of \textit{bifurcations} 
\cite{Scheffer2009, Scheffer2012, Jusup2022, Grziwotz2023}.
A bifurcation is a point at which a small change in a system's control parameter 
causes a qualitative shift in its equilibrium state, resulting in a new and distinct configuration of system behavior.
There are numerous examples of systems that display dynamics resembling those of bifurcations.
For instance in ecology, 
predator-prey dynamics can lead to the sudden extinctions of a species \cite{Hossain2022}, 
small shifts in climate patterns may induce interglacial periods \cite{lenton2008}, 
vegetation loss above a critical threshold results in desertification \cite{Hirota2011}, 
or nutrient overload can cause a lake to suddenly transition from a clear-water to a algae dominated state \cite{Carpenter1997}. 
In social systems, notable examples are 
financial markets experiencing long periods of low volatility followed by abrupt crashes \cite{Sorcausm18}
and 
growing dissatisfaction with political conditions, which can trigger widespread unrest or revolution \cite{Granovetter1978}. 
Health-related examples involve the sudden onset of conditions like 
microbiome dysregulation \cite{Trefois2015},  transitions between different states representing distinct illnesses based on varying internal and external influences \cite{Sorcobbsch09},
depressive episodes \cite{Van2014}, 
or epileptic seizures \cite{Maturana2020}.

As a system approaches a critical bifurcation threshold, it is often marked by increased variance and critical slowing down 
revealed, for instance, by increased autocorrelation in observable metrics. 
These statistical signatures reveal that the system's recovery from perturbations is slowing as it approaches a bifurcation, alongside an increasing sensitivity to fluctuations in the system's state. These patterns are frequently
 considered as early-warning signals for an impending bifurcation \cite{Scheffer2009, Scheffer2012, Jusup2022, Grziwotz2023}.

Although bifurcation theory provides valuable insights into the qualitative behavior of dynamical systems, it faces at least two major challenges.
First, it relies on the existence of a well-defined control parameter and its closeness to a critical threshold. Second, it assumes the presence of a driving force that pushes the system toward or beyond this threshold. These assumptions are not only difficult to verify empirically but also conceptually limiting, potentially undermining their broader applicability. It is, therefore, not surprising that the relevance of critical bifurcations for explaining observed system behaviors continues to be a topic of active debate. Even when such bifurcations exist somewhere in the control parameter space, they may not govern the system's actual dynamics. For instance, evidence remains mixed on whether the onset of epilepsy is genuinely driven by critical transitions \cite{Wilkat2019, Maturana2020}. Similarly, it is still unclear whether elements of the Earth system -- such as the Amazon rainforest or the Atlantic Meridional Overturning Circulation (AMOC) -- are nearing bifurcation points \cite{Ben2024}. 
While  our analysis does not dispute the existence of bifurcation points in these systems, we argue that the estimated proximity to such points may be significantly altered when non-normal dynamics, as described below, are taken into account.

We present a more general mechanism that applies to systems with a stable equilibrium attractor, highlighting how linear stochastic dynamics with non-normal structure around the attractor can transiently amplify perturbations in a distinctive way. We demonstrate that small perturbations around the equilibrium produce transient repulsions,
which we term \textit{pseudo-bifurcations}, with similar statistical signatures to those observed near genuine bifurcations. The terminology \textit{pseudo-bifurcations} draws from the mathematical foundation of non-normal operators, where positive pseudospectral eigenvalues characterize transient instabilities even in asymptotically stable systems.
Much like the effective dimension reduction seen during the onset of a bifurcation, we find that non-normal transients are primarily governed by dynamics confined to a two-dimensional subspace, where most of the transient amplification occurs.
We further find that these transients are accompanied by the same statistical markers mentioned above
(increased variance and autocorrelation). We demonstrate that, as a system approaches an actual bifurcation, non-normal transients systematically emerge beforehand, making it difficult to distinguish between the two phenomena and potentially introducing a bias that makes the system appear closer to criticality than it is.
Finally, we show that the onset of non-normal transients can be anticipated by measuring the system's alignment along its \textit{non-normal mode}, and provide a numerical algorithm for stable identification of these non-normal modes.
We empirically illustrate our findings by analyzing EEG signals that capture brain activity during epileptic seizures, revealing clear evidence of non-normal dynamics.
This opens a new path towards the anticipation and control of instabilities in complex systems.

Our work builds on prior extensions of critical phenomena theory to non-normal systems, particularly those characterized by non-orthogonal eigenbases in linearized operators \cite{Farrell1988,Farrell1989}. These ideas have been developed into a general framework for analyzing transient amplification in hydrodynamic flows and turbulent jets, particularly in the context of atmospheric instability \cite{FarrellIoannou1996a,FarrellIoannou1996b,FarrellIoannou2003,FarrellIoannou2007}. 
It has been shown that non-normality in such systems can lead to transient growth that closely resembles the behavior near bifurcations. 
We build on and generalize this perspective by applying it to a broader class of dynamical systems, well beyond the context of hydrodynamic flows. 
Specifically, our concept of pseudo-bifurcations captures the transient, bifurcation-like dynamics that emerge in non-normal systems. We further propose a method to distinguish between spectral and non-normal sources of early-warning signals. In addition, we identify new observables that improve the diagnostic utility of these signals. Together, these contributions expand the domain of early-warning diagnostics to a wider class of dynamical systems.

The only requirement to observe such bifurcation-like dynamics is that the system exhibits sufficient non-normality, meaning that the interactions among components are both sufficiently asymmetric and structured with a certain degree of hierarchy. These features are commonly found in many natural and social systems \cite{Asllani2018,OBrien2021}. 
These characteristics are not limited to networks. They also appear in systems like turbulent flows,
where non-normal dynamics arise from the interplay of viscous and advective terms,
which leads to asymmetric, non-orthogonal interactions among velocity perturbations \cite{Trefethen1993}.  
In models of mechanical deformation under shear, such as those describing systems of interacting blocks or grains in fault gouge, the governing equations exhibit tensorial rotational mechanisms that are governed by non-normal dynamical operators. These can lead to substantial transient amplification of small perturbations, potentially acting as a triggering process for seismic events \cite{Charan2020,Charan2021Remote}.

Our findings are generic.
To observe a pseudo-bifurcation, all one needs to assume is that the Jacobian of the linearized system,
around its equilibrium point, forms a non-normal matrix. This is a very general condition since non-normal matrices form a measure-one set in the space of all matrices
and  because hierarchical asymmetric interactions are common across physical and social sciences.
Our findings suggest that many systems previously interpreted as approaching criticality may warrant re-evaluation, as they could instead represent asymptotically stable yet non-normal systems exhibiting pronounced transient instabilities.
More generally, our results offer a different perspective on how to interpret precursory signals in dynamical systems and highlight the role of non-normal dynamics in improving our understanding and management of a wide range of physical,
natural, and social systems.

\section*{Mathematical Foundation}

To lay the mathematical foundation of our framework,
we now introduce the class of stochastic non-normal linear systems that forms the core of this study.

\begin{definition}  \label{def:snl}
A $N$-dimensional stochastic non-normal linear system (SNL) is defined by
\begin{equation}    \label{eq:basic_dynamics}
\dot{\x}_t ~=~ \A~\x_t + \sqrt{2\delta}~\et_t
,\quad \et_t~\overset{\textit{i.i.d}}{\sim}~\mathcal{N}(0,\I).
\end{equation}
We restrict our analysis to the cases where $\A$ is a full-rank $N \times N$ diagonalizable negative-definite matrix that is non-normal,
i.e. it cannot be diagonalized by a unitary transformation, although it can still be diagonalized by a non-unitary one.
Non-normality also means that $A A^\dag \neq A^\dag A$. The positive paramter $\delta$ is the variance of the noise.
\end{definition}

Qualitatively, the non-normality of $\A$ reflects inherently directional and hierarchically structured interactions \cite{Asllani2018,OBrien2021}.
Many systems across the natural and social sciences exhibit non-normality,
including turbulence in fluid dynamics \cite{Trefethen1993},
perturbations in ecosystems \cite{Neubert1997, Tang2014},
non-Hermitian quantum mechanics \cite{Hatano1996},
population dynamics \cite{Neubert2002},
synchronization of optoelectronic oscillators \cite{Ravoori2011},
amplification in neural activity \cite{Murphy2009},
chemical reactions \cite{Nicoletti2019},
network synchronisation \cite{Asllani2018b, Nicoletti2018, Muolo2020},
neuronal networks \cite{Hennequin2012, Gudowska2020},
and financial markets \cite{Sornette2023, Wang2024}.
This is unsurprising, since non-normal matrices, and by extension non-normal dynamics, are the rule rather than the exception.
Mathematically, non-normal matrices form a measure-one set in the space of all matrices,
while normal matrices form a measure-zero subset.
This prevalence implies that most real-world systems are inherently non-normal
and therefore prone to characteristic behaviors, such as transient growth and heightened sensitivity to perturbations that we detail below.

We work with diagonalizable, non-normal matrices $\A$.
Let $\PP$ denote the basis transformation that diagonalizes $\A$, i.e., $\A = \PP\Lam\PP^{-1}$,
where $\Lam = \text{Diag}(\lambda_i,|, i = 1, \cdots, N)$.
Without loss of generality, we perform a singular value decomposition (SVD) of $\PP$ as
$\PP = \U\Sig\V^\dag$,
where $\U$ and $\V$ are unitary matrices, and $\Sig = \text{Diag}(\sigma_i,|, i = 1, \cdots, N)$
contains the singular values of $\PP$, assumed to be ordered such that $\sigma_i \ge \sigma_{i+1}$.
The condition number $\kappa = \kappa(\PP)$ of matrix $\PP$ is defined as 
the ratio between the largest and smallest singular values:
$\kappa = \sigma_1 / \sigma_N$.
If $\A$ is normal, then $\PP$ is unitary and $\kappa = 1$ (all singular values are equal to $1$).
In contrast, for non-normal systems, $\kappa > 1$.
Non-normality implies that at least two eigenvectors of $\A$ are non-orthogonal, leading to a poorly conditioned eigenvector matrix $\PP$. 
As this non-orthogonality increases and the eigenvectors begin to coalesce, $\PP$ approaches singularity and  $\kappa \to \infty$. 
Thus, $\kappa$ serves as a quantitative measure of the degree of non-normality.

\begin{definition}  \label{def:deg_nn}
A degree of non-normality is any scalar quantity that increases monotonically as a matrix deviates further from normality.
\end{definition}
A degree of non-normality quantifies ``how non-normal'' a matrix is, in the sense that it captures the extent to which the eigenvectors of the matrix fail to be orthogonal,
or the degree to which transient amplification is possible even in asymptotically stable systems.
There exist multiple ways to measure non-normality,
each emphasizing different facets of the phenomenon.
For example, Henrici's departure from normality quantifies the average non-orthogonality of the eigenvectors via the Frobenius norm of the commutator $[\A, \A^\dag]$,
while the Kreiss constant bounds the worst-case transient amplification induced by a stable operator.
Other measures -- such as the pseudospectral abscissa, the numerical abscissa,
or the angle between left and right eigenvectors -- offer additional insights into sensitivity, short-term growth, and geometric structure.

In this paper, we focus on the condition number $\kappa(\PP)$ of the matrix $\PP$ that diagonalizes $\A$. We use $\kappa(\PP)$ as a quantitative measure of non-normality, suited to our comparison between genuine bifurcations and pseudo-bifurcations in non-normal systems.
The condition number captures the maximal distortion of the eigenbasis under the action of $\A$ and is particularly useful in our setting for three reasons.
First, it directly controls the strength of the transient response in the reduced two-dimensional dynamics (see \propref{pr:dim_red} below).
Second, it enables us to define a precise threshold $\kappa_c$ beyond which transient repulsion emerges (see \defref{def:critical_kappa} below).
Finally, it serves as a meaningful indicator of when the dimensionality reduction remains valid (see \propref{pr:dim_red} below):
as $\kappa$ increases, the dynamics increasingly concentrate along the non-normal and reaction modes, 
supporting a lower-dimensional representation of the system's behavior.

\section*{Overview of Bifurcation Theory and Early-Warning Signals}

\begin{definition}  \label{def:bifurcation}
Consider a dynamical system of the form $\dot{\x}_t = \f(\x_t)$,
where $\f$ is a smooth vector field with a stable fixed point $\x^*$, i.e. $\f(\x^*)=0$.
Linearizing the system around $\x^*$ yields
$
\dot{\x}_t = \J_f(\x_t - \x^*) + \mathcal{O}((\x - \x^*)^2),  
$
where $\J_f = \mathbf{D}\f|_{\x=\x^*}$ is the Jacobian at the fixed point and is negative definite, meaning all the real parts of its spectrum are negative.
The system is said to be near a bifurcation if for at least one eigenvalue $\lambda \in \sigma(\J_f)$,
$\text{Re}(\lambda) \to 0^-$. From here onward, we denote by $\lambda$ the eigenvalue of $\A$ with the largest real part.
\end{definition}

A bifurcation in a continuous dynamical system occurs when a small change in a parameter leads to a qualitative change in the stability of an equilibrium or fixed point.
This corresponds to the real part of an eigenvalue crossing zero (\defref{def:bifurcation}). Conversely, a system is said to approach a bifurcation
when the real part of the leading eigenvalue approaches zero from below as defined in (\defref{def:bifurcation}).
If the eigenvalue has a nonzero imaginary part, the bifurcation is classified as a Hopf bifurcation, leading to periodic limit cycles.
Otherwise, it is a steady-state bifurcation, such as a fold bifurcation (where two equilibria collide),
a pitchfork bifurcation (where one fixed point branches into three),
or a transcritical bifurcation (where two fixed points exchange stability).
Bifurcations have significant real-world consequences.
For instance, desertification in semi-arid regions like the Sahel occurs when vegetation loss exceeds a critical threshold,
triggering a rapid and often irreversible shift to desert landscapes \cite{Hirota2011}.
Similarly, eutrophication in lakes such as Lake Erie involves a bifurcation
where nutrient overload causes a sudden shift from clear water to an algae-dominated state \cite{Carpenter1997}.

The following definition requires specifying the ensemble average in the long-time limit as
\begin{equation}
    \mw{x} ~=~ \lim_{t\to\infty}~ \mw{x_t}~,
\end{equation}
where $\mw{x_t}$ denotes the statistical ensemble average of $x_t$ (i.e. the average over the noise realisations $\eta_t$).
The autocorrelation function is 
\begin{equation}
    C(\tau) ~=~ \lim_{t\to\infty}~\frac{\mw{x_t x_{t+\tau}}}{\mw{x_t^2}} .
    \label{hbgwqbq}
\end{equation}

\begin{definition}  \label{def:early_warning}
Close to a bifurcation (\defref{def:bifurcation}), the system's dynamics can be effectively captured by a one-dimensional normal form 
$\dot{x}_t = f(x_t)$, a process known as \textit{dimensional reduction} where the dominant behavior unfolds along 
a single direction associated with the vanishing eigenvalue. 
A common approximation assumes that the system follows a linearised Ornstein-Uhlenbeck (OU) process around
the fixed-point $x^*$:
\begin{equation}    \label{eq:ou}
\dot{x}_t ~=~ \lambda~(x_t - x^*) + \sqrt{2\delta}~\eta_t,
\quad \eta_t \overset{\textit{i.i.d}}{\sim} \mathcal{N}(0,1),
\end{equation}
where $\f(x^*) = 0$ and $\lambda = f'(x^*)$.
As the system approaches a bifurcation, the following early-warning signals are expected:
\begin{itemize}
\item[E1] The approach toward a bifurcation in continuous time is characterized by the real part of the leading eigenvalue $\lambda$ approaching zero from below, i.e., $\lambda \to 0^-$ in equation (\ref{eq:ou}). In discrete-time systems, this corresponds to the spectral radius of the Jacobian approaching one from below, a regime commonly referred to as the ``unit root'' limit.
\item[E2] Progressive increase in variance expressing enhanced fluctuations, i.e. $\mw{x^2} \to \infty$ as $\lambda \to 0^-$.
\item[E3] Critical slowing down: as $\lambda \to 0^-$, the characteristic autocorrelation decay time increases,
i.e. $\tau_0 = \left[-\tfrac{d}{d\tau}\ln (C(\tau))|_{\tau=0}\right]^{-1} \to \infty$.
\end{itemize}
\end{definition}
These early-warning signals {\it E1-E3} are widely used to anticipate impending bifurcations or critical transitions in dynamical systems.
They are particularly valuable in complex systems, ecological, socioeconomic, and climatic, where they help identify critical thresholds beyond which qualitative shifts in system behaviour occur \cite{Scheffer2009, Scheffer2012, Jusup2022, Grziwotz2023}.
These signatures reflect reduced system resilience and slower recovery from perturbations as bifurcation nears (\defref{def:early_warning}),
enabling timely interventions and improved management strategies.

\section*{Results}

The primary objective of this paper is to establish the following propositions {\bf 1-3}
both analytically and numerically, which characterize the emergence of early-warning signals in stochastic non-normal systems. 
\begin{proposition} \label{pr:es_snl}
    All early-warning signals associated with bifurcations (\defref{def:early_warning}) also emerge transiently in an SNL 
    (\defref{def:snl}) far from bifurcation, even when the matrix $\A$ remains fixed. 
\end{proposition}
We note that, according to \defref{def:snl}, matrix $\A$ is non-normal. And it being fixed means, in particular, that is spectrum remains stable. 
The following proposition extends these insights to situations in which $\A$ itself is changing.
\begin{proposition} \label{pr:es_snl_not_fixed}
    The early warning signals described in \propref{pr:es_snl} are particularly pronounced and more persistent when the matrix $\A$ has a fixed, stable spectrum but a time-varying degree of non-normality (\defref{def:deg_nn}). In this case, the signals closely mimick the dynamics of a genuine bifurcation.
\end{proposition}
While \propref{pr:es_snl} and \propref{pr:es_snl_not_fixed} establish that the early warning systems previously attributed to bifurcations also emerge in SNL systems, the following proposition further establishes that the former induces the later.
\begin{proposition} \label{pr:pseudo_bifurcation}
    All $N$-dimensional systems (with $N>1$) enter a pseudo-bifurcation regime (elaborated below) 
    before reaching a bifurcation.
     Pseudo-bifurcations thus emerge as a necessary precursor to bifurcations, positioning genuine bifurcations as a special, limiting case within the broader class of pseudo-bifurcation phenomena.
\end{proposition}

After establishing \propref{pr:es_snl}-\propref{pr:pseudo_bifurcation}, we demonstrate their empirical relevance using EEG recordings from patients experiencing epileptic seizures. Our findings reveal that traditional early-warning signals can arise purely from non-normal dynamics -- even when the system's spectrum remains unchanged -- thereby complicating their interpretation and emphasizing the need for complementary observables to reliably distinguish true bifurcations from pseudo-critical behavior.

These results highlight a central challenge in the early-warning signal and complex systems literature: distinguishing genuine signals of approaching bifurcations from spurious patterns. Indicators such as rising variance, increasing autocorrelation, and drift toward a unit root are known to be highly sensitive to methodological choices, including detrending methods and the size of rolling windows used in analysis \cite{Boettiger2012,Dakos2012,Drake2010}. This sensitivity undermines the robustness of such indicators and raises the risk of false positives, not merely due to statistical noise, but more fundamentally because similar signal patterns can emerge in the absence of any true bifurcation. In particular, non-normal dynamics can produce early-warning-like behavior even when the system remains far from a critical threshold. Thus, the issue is not just one of statistical significance, but of structural misinterpretation: apparent early-warning signals may reflect complex transient dynamics rather than genuine critical transitions.

\subsection*{Reduced Non-Normal Dynamics}

While \defref{def:snl} defines general stochastic non-normal systems, the key features of transient amplification manifested as finite-lived deviations driven by the non-normal structure of the matrix $\A$ \eqref{eq:basic_dynamics}  can in practice be effectively captured within a two-dimensional subspace. Our aim is to construct a minimal two-dimensional model that can describe two situations so as to compare them in their precursory behaviors: (i) proximity to instability reflected by $\lambda := \max \text{Re}(\sigma(\A)) \to 0^-$ and (ii) pronounced non-normality quantified by a large condition number $\kappa \gg 1$. 
We now describe how to systematically reduce the full system to its dominant 
non-normal and reactive modes.

\begin{definition}  \label{def:nn_1nm}
A non-normal matrix $\A$ is said to exhibit a unique non-normal mode  
if the singular values of its eigenbasis transformation matrix $\PP$
satisfy $\sigma_1 \approx \sigma_2 \approx \cdots \approx \sigma_{N-1} \approx 1$,
but the smallest singular value is much smaller, i.e., $\sigma_N \ll 1$
such that $\kappa = \sigma_1/\sigma_N\gg 1$.
The corresponding non-normal mode is defined as the column $\hn \sim \hu_N$
of matrix $\U$, i.e. without loss of generality the last column of $\U$
 in the SVD $\PP = \U \Sig \V^\dag$
associated with the smallest singular value $\sigma_N$.
\end{definition}

Using this singular value structure,
we derive a remarkably simple two-dimensional system
that captures the essence of non-normal transient growth.

\begin{proposition} \label{pr:dim_red}
For a matrix satisfying \defref{def:nn_1nm},
we can, up to a unitary transformation, reduce the system to the two-dimensional dynamics:
\begin{subequations}    \label{eq:dim_red}
    \begin{align}
        &\dot{n}_t = -\alpha n_t + \kappa^{-1} r_t + \sqrt{2\delta}~\eta_{n,t} \label{eq:n} \\
        &\dot{r}_t = \kappa n_t - \alpha r_t + \sqrt{2\delta}~\eta_{r,t} \label{eq:r}
    \end{align}
\end{subequations}
Here, $n$ denotes the projection along the non-normal mode $\hn = \hu_N$,
and $r$ is an orthogonal projection referred to as the reaction mode,
defined by $\hr \sim \sum_{i=1}^{N-1} \hu_i$. Here $\alpha$ is the characteristic relaxation rate 
towards the stable fixed point.
Note that the eigenvalues of this system are $\lambda_\pm = -\alpha \pm 1$.
Since, by assumption, the equilibrium state is stable, it follows that $\alpha > 1$, meaning the dominant eigenvalue satisfies $\lambda_+ < 0$.
\end{proposition}

\begin{figure}
    \centering
    \includegraphics[width=0.9\linewidth]{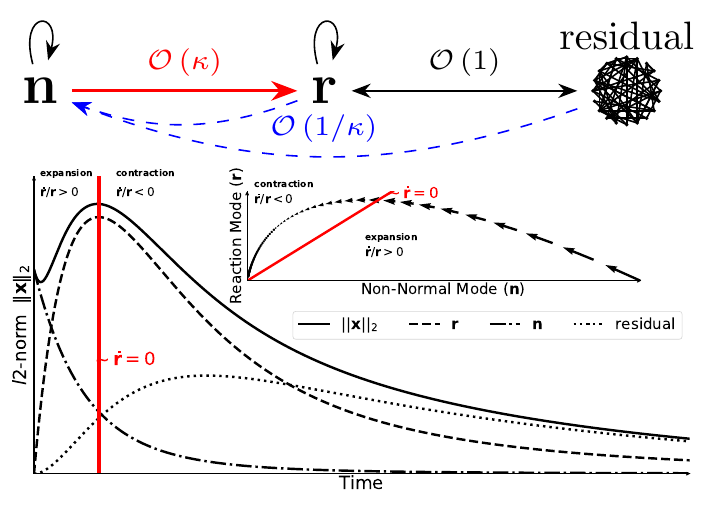}
    \caption{
    (top)
    Schematic representation of equation \eqref{eq:basic_dynamics} in its reduced form \eqref{eq:dim_red}.
    (bottom)
    Associated transient dynamics for $||\x||_2$ both as a function of time (main) and in the reduced $(n,r)$-plane (inset). 
    }
    \label{fig:confied_dynamics}
\end{figure}

An illustration of the dimensionality reduction described in \propref{pr:dim_red} is shown in Figure \ref{fig:confied_dynamics} (top).
In this representation, most interactions in the original $N$-dimensional system are $\mathcal{O}(1)$,
except for the coupling from the non-normal mode to the reaction mode, which scales as $\mathcal{O}(\kappa)$,
and the reverse coupling from the reaction mode to the non-normal mode, which is of order $\mathcal{O}(\kappa^{-1})$.
When the system's degree of non-normality is sufficiently high,
the dynamics described by \eqref{eq:basic_dynamics} exhibit pronounced transient repulsion.
To understand this, consider the deterministic form of \eqref{eq:dim_red}, omitting the stochastic and higher-order terms.
Assuming an initial condition $(n, r) = (1, 0)$,
the variable $n$ (and hence the norm $||\mathbf{x}||_2$) initially decreases as the non-normal mode relaxes at rate $\alpha$ via \eqref{eq:n}.
Meanwhile, the reaction component $r$ grows rapidly at rate $\kappa$ via \eqref{eq:r}.
For sufficiently large $\kappa$, this transient growth outweighs the relaxation,
causing $||\mathbf{x}||_2$ to temporarily increase.
The peak is reached when $\dot{r} = 0$,
after which the system decays back to equilibrium at rate $|\lambda_+|$
(see Figure \ref{fig:confied_dynamics}, bottom).
This two-dimensional decomposition holds regardless of the original system dimension $N \geq 2$
as shown in the Supplementary Material.

For the remainder of the paper, we use the reduced non-normal system \eqref{eq:dim_red}
as the basis for our mathematical derivations,
as it fully encapsulates the key transient features of non-normal dynamics.

\subsection*{Characterization and Prediction of Non-Normal Transients}

We now turn to a systematic analysis of the dynamical signatures produced by non-normal systems,
focusing on how transient responses can reproduce early-warning signals (\defref{def:early_warning}),
when we assume the dynamics to be in its reduced non-normal form (\propref{pr:dim_red}).

\begin{figure}[!htb]
    \centering
    \includegraphics[width=\linewidth]{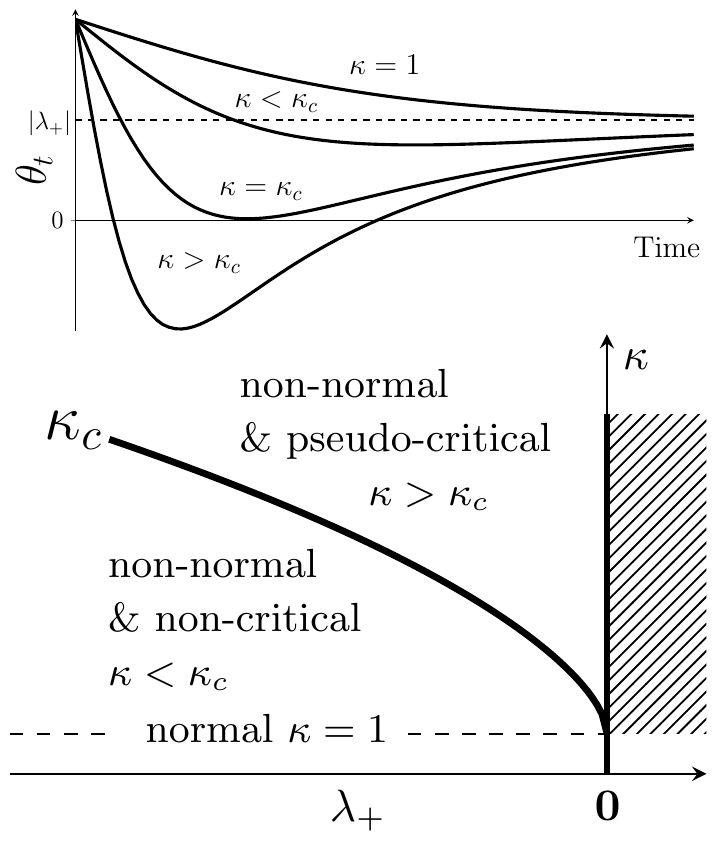}
    \caption{
    (Top) 
    Behavior of the Ornstein-Uhlenbeck parameter $\theta_t$ (\ref{eq:theta_root})    
    for different levels of non-normality $\kappa$. 
    For $\kappa > \kappa_c$, the typically positive mean-reversion coefficient $\theta_t$ transiently becomes negative, thereby inducing repulsive growth
    away from the stable fixed point.
    (Bottom)
    Phase diagram illustrating the system's regimes as a function of the leading eigenvalue $\lambda_+=-\alpha +1$ 
    from the reduced model~\eqref{eq:dim_red} and the condition number $\kappa$, which quantifies non-normality.
    As the bifurcation at $\lambda_+ = 0$ is approached at fixed $\kappa$, the system first traverses the pseudo-critical regime as $\kappa_c$ becomes
    smaller than $\kappa$. 
    }
    \label{fig:regimes}
\end{figure}

\begin{proposition} \label{pr:effect_ou}
    In order to characterize the parameter regime in which the non-normal transients dominate the dynamics,
   equation \eqref{eq:r} can be rewritten as an effective Ornstein-Uhlenbeck (OU) process (analogous to an AR(1) process in discrete time):
    \begin{equation} \label{eq:r_reduced}
        \dot{r}_t ~=~ - \theta_t ~r_t + \sqrt{2\delta} ~\eta_{r,t},
    \end{equation}
    where $\theta_t$ is a time-dependent mean-reversion rate with memory given by (see Methods) 
    \begin{equation}\label{eq:theta_root}
  \theta_t ~=~
  \alpha
  -\frac{(\kappa^{2}+1)\cosh t~\sinh t}{\kappa^{2}\sinh(t)^{2}+\cosh(t)^{2}}~.
\end{equation}
 The upper graph of Figure \ref{fig:regimes} shows that $\theta_t$ decays to $0$ at long times.
\end{proposition}
A direct consequence of this time-dependent mean-reversion (\ref{eq:theta_root}) is the emergence of a well-defined non-normality threshold that delineates the onset of repulsive transient dynamics.
\begin{proposition}  \label{def:critical_kappa}
    Based on the representation of the reaction as an effective Ornstein-Uhlenbeck process (\ref{eq:r_reduced}) with (\ref{eq:theta_root}), 
    a critical condition number $\kappa_c$ naturally emerges from the analysis:
\begin{equation} \label{eq:kappa_critical}
    \kappa_c = \alpha + \sqrt{\alpha^2 - 1}~.
\end{equation}
For $\kappa \ge \kappa_c$, there exists a first time $t^*>0$ at which $\theta_t$
reaches zero, after which it becomes negative over a finite time interval $(t^*, t_>)$ for some $t_> > t^*$.
This corresponds to a finite time interval during which the reaction $r$ is transiently pushed away from its stable equilibrium $r = 0$
(see Figure \ref{fig:confied_dynamics} top).
The system thus exhibits local instability despite its asymptotic stability.
\end{proposition}

For systems with $\kappa > \kappa_c$, the stochastic dynamics of $\|\x\|_2$ is characterized by intermittent shifts between two regimes.
For relatively small values of $\|\x\|_2$, noise dominates,
and the behavior appears primarily stochastic around the equilibrium point.
If, by chance, the system's trajectory becomes increasingly aligned with the non-normal mode,
transient amplification takes hold and the system is dominated by the approximately deterministic transient repulsion,
replicating the early-warning signal \textit{E1} in \defref{def:early_warning}. 
This follows from the observation that $\theta_t$ crosses zero at a finite time, thereby emulating the signature of a true bifurcation.
It is important to note that this early-warning signal \textit{E1} emerges spontaneously and persists for a finite duration,
despite the system being held at a constant distance from the bifurcation point and a fixed level of non-normality $\kappa$. 

Moreover, this enables the integration of criticality and non-normality within a unified framework,
since when $\lambda_+\to 0^-$ i.e. $\alpha\to 1^+$; the critical condition number tends to one,
and so for all $\kappa>\kappa_c\approx 1$, the system is dominated by non-normal transients.
Therefore, all systems tend to exhibit pseudo-critical behavior driven by non-normal dynamics 
before reaching a truly critical state characterized by spectral changes. 
This progression is illustrated in Figure~\ref{fig:regimes} (bottom), which presents a phase diagram in the  $(\lambda_+,\kappa)$ plane. 
The diagram delineates non-critical, pseudo-critical, and critical regimes within a unified framework and 
highlights their respective implications for the effective time-dependent mean-reversion rate $\theta_t$
 as defined in equation~\eqref{eq:r_reduced}.
 
Building on this framework, we now establish in \propref{pr:variance_reaction} that the variance along the reaction direction is given by an explicit expression that captures its dependence on both the proximity to criticality and the degree of non-normality $\kappa$.
\begin{proposition} \label{pr:variance_reaction}
    The variance along the reaction direction is given by
     \begin{equation}    \label{eq:variance_reaction}
        \mw{r^2} ~=~ \lim_{t\to\infty}~ \mw{r_t^2} ~=~ \frac{\delta\alpha}{\alpha^2 - 1}\frac{\kappa^2 + 1}{2}
        \left[1 - \frac{\kappa^2 - 1}{\kappa^2 + 1}\frac{\alpha^2 - 1}{\alpha^2}\right].
    \end{equation}
    and thus scales asymptotically as $\mathcal{O}(\kappa^2)$, which dominates the total variance of the system variable.
\end{proposition}
\propref{pr:variance_reaction} shows that a non-normal system is able to reproduce the early-warning signal  \textit{E2} 
in \defref{def:early_warning} of the increase in variance 
in systems held at a constant distance from the bifurcation point in the presence of an increasing degree of non-normality $\kappa$. 

Furthermore, the following proposition shows that the instantaneous correlation time $ \tau_0$
satisfies a specific relation, revealing a local slowing-down effect that becomes more pronounced as $\kappa$ increases,
a phenomenon we refer to as non-normal slowing down.
\begin{proposition} \label{pr:slowing_down}
   Given that the reaction is described by \eqref{eq:r}, its instantaneous correlation time $\tau_0$ satisfies
    \begin{equation}
        \tau_0 = \left[-\left.\frac{d}{d\tau}\ln C\right|_{\tau=0}\right]^{-1}
        ~=~ \frac{\mw{r^2}}{\delta} ~=~ \mathcal{O}(\kappa^2)~,
    \end{equation}
    where the auto-correlation $C(\tau)$ along the reaction is defined by (\ref{hbgwqbq}). This indicates that
    the system is locally slowing down. The correlation time tends to infinity for \(\kappa\to\infty\).
    We refer to this phenomenon as the ``non-normal slowing down effect''.
\end{proposition}
\propref{pr:slowing_down} shows that a non-normal system is able to reproduce the early-warning signal  \textit{E3} 
of critical slowing down, even when held at a constant distance from the bifurcation point in the presence of an increasing degree of non-normality $\kappa$. 

\subsection*{Quasi-Deterministic Dynamics and Cycles}

\begin{figure*}[!htb]
    \centering
    \includegraphics[width=1.05\linewidth]{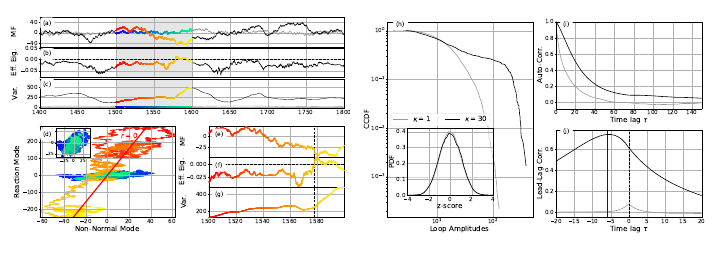}
    \caption{
    (a-c)
    Mean-field variable, effective eigenvalue $\lambda_\text{eff}$ and variance defined via \eqref{eq:MF_AR1} from a simulated system \eqref{eq:basic_dynamics} with $N=10$ dimensions:  normal case (grey line, $\kappa=1$) and non-normal case (black line, $\kappa=30$) with otherwise identical conditions. 
    (e-g)
    Same as (a-c) but only for the non-normal system for a specific excerpt for better resolution.
    The vertical dashed line highlights a pseudo-bifurcation at which the effective eigenvalue crosses zero. 
    (d)
    Reduced dynamics \eqref{eq:dim_red} of \propref{pr:dim_red} in the $(n,r)$-plane. 
    While the non-normal system (yellow to red) displays distinct transient repulsion and contraction along the reaction mode $r$, the normal system (green to blue) shows no
     such trends (as also highlighted in the re-scaled inset figure).    
    (h)
    Complementary cumulative distribution (CCD) of projection along the reaction mode $r$ from subplot (d).
    The CCD of the non-normal system (dark line) is a mixture of two distributions: one dominated by noise, and the other dominated by quasi-deterministic loops. 
    The second distribution is absent in the normal case (grey line). 
    The probability distribution function (PDF) of the mean-field (inset) shows no such distinction between the normal and non-normal systems. 
    (i)
    Autocorrelation of the mean-field dynamics. 
    (j)
    Lead-lag correlation between the system's alignment along the non-normal mode and the mean-field dynamics, with a clear indication that the former leads the later when the system is non-normal.
    }
    \label{fig:simulations}
\end{figure*}

To formalize the notion of transient predictability observed in non-normal systems, we introduce the concept of quasi-deterministic dynamics, which characterizes regimes where local predictability persists over finite time scales despite the presence of noise.
\begin{definition}  \label{def:quasi_deterministic}
   A dynamic is considered quasi-deterministic when its autocorrelation remains close to one over a finite time duration, 
   enabling short-term local predictability and giving rise to transient pockets of predictability.
\end{definition}
This concept of quasi-deterministic dynamics plays a central role in distinguishing non-normal dynamics from true bifurcations, as the origins of early-warning signals (\defref{def:early_warning}) differ fundamentally in the two cases.
In bifurcating systems, early-warning signals stem from changes in the spectral properties of the linearized dynamics
as the control parameter is varying.
In contrast, for non-normal systems, these signals originate from the pseudo-spectrum of the system's linear operator, that is, from how the system responds to small perturbations beyond what is predicted by its eigenvalues alone \cite{Embree2005}.
These signals thus reflect the degree of non-normality (\defref{def:deg_nn}), for instance as quantified by the condition number $\kappa$ of the eigenbasis transformation.
Moreover, the early-warning signal {\it E1} emerges spontaneously despite the non-normal system being held at a constant distance 
from the bifurcation point and a fixed level of non-normality $\kappa$. 
The other two early-warning signals {\it E2} and {\it E3} arise in a mathematically well-defined manner as the degree of non-normality $\kappa$
increases, even when the system remains at a fixed distance from the bifurcation point. These signals can also appear transiently as emergent behaviors at constant $\kappa$.

\begin{proposition}  \label{def:stochastic_loop}
In a stochastic non-normal linear (SNL) system as defined in \defref{def:snl}, with a single non-normal mode (\defref{def:nn_1nm}) and strong non-normality 
($\kappa \gg 1$), the non-normal slowing down effect (\propref{pr:slowing_down}) implies that deviations along the non-normal mode evolve quasi-deterministically (\defref{def:quasi_deterministic}) over a finite time interval. As a consequence, the system exhibits quasi-deterministic cycles in the phase space spanned by the non-normal and reaction modes: small stochastic perturbations trigger coherent excursions that follow quasi-deterministic trajectories before relaxing back to the stable fixed point. The stochasticity affects primarily the timing and amplitude of the initial deviation, while the subsequent evolution is largely predictable.
\end{proposition}
Quasi-deterministic loops that emerge from stochastic fluctuations are a hallmark of early-warning signals driven by non-normality, rather than by bifurcations.
In SNL systems, they serve as the analogue of limit cycles observed in Hopf bifurcations, though arising without a change in spectral stability.
\newline

\section*{Pseudo-Bifurcations}

All of the properties described above can be traced back to the fact that non-normal dynamics induce transient, repulsive, and quasi-deterministic growth, as formalized in \eqref{eq:r} and defined in \defref{def:quasi_deterministic}. This growth leads to coherent excursions in the phase space $(n,r)$ spanned by the non-normal and reaction modes, which take the form of quasi-deterministic loops (\propref{def:stochastic_loop}). Qualitatively, these loops resemble Hopf-like limit cycles, as illustrated in Figure~\ref{fig:simulations}(d) below.

This observation highlights a key insight: non-normality, much like genuine bifurcations, can produce early-warning signals such as increased variance, autocorrelation, and slowing down. However, these signals arise purely from transient amplification, without any actual change in the system's spectral stability. As a result, they can be mistaken for signs of an imminent critical transition, thereby complicating the interpretation of early-warning indicators.

To describe this phenomenon, we introduce the concept of a pseudo-bifurcation defined as a transient bifurcation-like regime driven not by spectral changes, but by non-normal amplification.
\begin{definition} \label{def:pseudo-bifurcation}
A pseudo-bifurcation is a transient bifurcation-like phenomenon driven by non-normal dynamics, rather than by changes in spectral stability.
\end{definition}

This terminology is further motivated by pseudospectrum theory \cite{Embree2005}, which naturally captures the transient behavior of non-normal systems. Unlike standard eigenvalue analysis, the pseudospectrum accounts for the amplification of perturbations that occur even when the eigenvalues indicate asymptotic stability. It thus provides the appropriate analytical framework for studying pseudo-bifurcations and distinguishing them from true critical transitions.

\subsection*{Numerical Illustration}

We illustrate these results by simulating an $N=10$ dimensional system with noise variance parameter $\delta=1/2$.
The matrix $\A$ has negative eigenvalues \(\lambda_1 = -10, \lambda_2 = -9, \ldots, \lambda_{10} = -1\),
forming the diagonal matrix \(\Lam = \text{diag}(\lambda_1, \ldots, \lambda_{10})\).
We further take the first \(N-1\) singular values \(\sigma_1, \ldots, \sigma_{N-1}\) of the transformation matrix \(\PP\) all equal to \(1\). 
This allows us to tune the system’s degree of non-normality by setting the last singular value \(\sigma_{10} = \kappa^{-1}\).
The matrix \(\A\) is then generated via \(\A = \PP \Lam \PP^{-1}\), where 
\(\PP= \mathbf{U} \Sig \mathbf{V}^\dagger\), with 
\(\mathbf{U}\) and \(\mathbf{V}^\dagger\) being two randomly generated orthonormal matrices,
and \(\Sig = \text{diag}(\sigma_1, \ldots, \sigma_{10})\).

Figure \ref{fig:simulations} shows two trajectories: one where \(\A\) is normal (\(\kappa=1\)) and one where it is non-normal (\(\kappa=30\)), under otherwise identical conditions. 
The system's mean-field value \(\bar{x}_t \equiv \frac{1}{N} \sum_{i=1}^N x_{i,t}\) is punctuated by large repulsive transients in the non-normal case compared to the normal one.
The quasi-deterministic, long-memory transient behaviour along the reaction mode \(r\) is evident in the complementary cumulative distribution of \(\|\x\|_2\),
which exhibits a mixture of two well-separated contributions: one for normal fluctuations and one associated with large stochastic loops.

Assuming full knowledge of \(\A\), an assumption we relax later, allows us to anticipate the onset of non-normal transients.
A useful predictor is the projection \(n\) of the system onto the non-normal mode \(\hn\), which governs the transient growth along \(\hr\) via \eqref{eq:r}. 
The lead-lag correlation between \(n\) and \(\bar{x}\) as a function of the lag \(\tau\) reveals that \(n\) significantly leads the onset of large transients (Figure \ref{fig:simulations} (j)) in non-normal systems,
whereas no such predictability is observed in the normal case.
This suggests that, by monitoring a system's alignment along the non-normal mode \(n\),
one can anticipate the emergence of a transient perturbation.

\subsection*{Non-Spectral Control Parameter $\kappa$ Versus Spectral Stability Control of the Distance to Bifurcation Point}

\propref{pr:es_snl} has been established by using analytical derivations and numerical simulations of a known system (\ref{def:snl}).
In reality, it is almost never the case that we have perfect observation of $\x$,
nor a clear understanding of the underlying dynamics. What is in general available
is the observation of the system’s mean-field state value $\bar{x}$ (or equivalently, its norm $\|\x\|_2$).
In the absence of further information, one can model the mean-field dynamics as an effective Ornstein-Uhlenbeck process,
\begin{equation} \label{eq:MF_AR1}
\dot{\bar{x}}_t ~=~ \lambda_\text{eff}~\bar{x}_t ~+~ \sqrt{2\delta_{\text{eff}}} ~\eta,
\end{equation}
where $\bar{x}$ corresponds to $x_t-x^*$ of equation (\ref{eq:ou})
and the effective eigenvalue $\lambda_{\text{eff}} < 0$ captures the effective mean-reverting component of the system.
The fluctuations of $\bar{x}_t $  have a variance $\delta_{\text{eff}} /|\lambda_{\text{eff}}|$,
and the correlation between states separated by a time interval $\Delta t$ is given by $\exp(- |\lambda_{\text{eff}}| \Delta t)$.
As $\lambda_{\text{eff}} \to 0^-$, both the variance and autocorrelation time diverge (see \defref{def:early_warning}), 
revealing increasing susceptibility and critical slowing down, respectively.
Figure \ref{fig:simulations} shows the evolution of $\lambda_\text{eff}$ and corresponding variance and auto-correlation
obtained by fitting an Ornstein-Uhlenbeck process to the mean-field dynamics 
of the non-normal system \eqref{eq:basic_dynamics}.
During large transients, $\lambda_\text{eff}$ approaches zero,
while the autocorrelation and variance increase, mimicking the approach to a critical bifurcation.

\begin{figure}[!htb]
    \centering
    \includegraphics[width=0.9\linewidth]{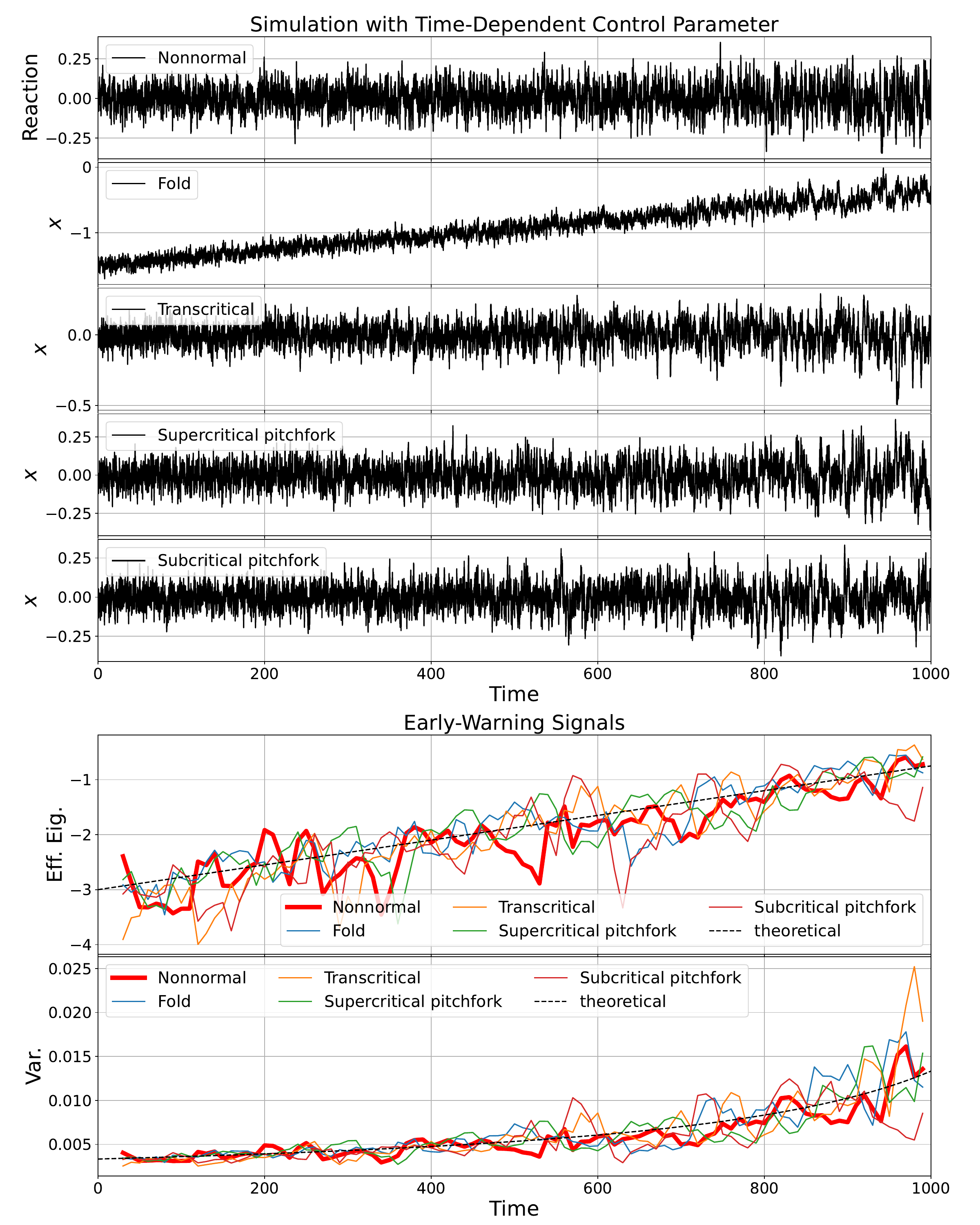}
    \caption{
        The five top panels show time series driven by time-dependent control parameters. For the non-normal system (Reaction, top), the control parameter is the condition number $\kappa$.
        For the normal forms (Fold, Transcritical, and Supercritical/Subcritical Pitchfork), the control parameter governs the system's approach to the bifurcation (see Supplementary Material for details).
        All systems are subjected to additive i.i.d. Gaussian noise with amplitude $\delta = 10^{-2}$.
        The two bottom panels display, over a rolling window of 30 time units, the evolution of the effective eigenvalue and the variance for each simulation.}
        \label{fig:early_warning}
\end{figure}

This resemblance to critical bifurcation behavior, evidenced by rising autocorrelation and variance, raises the question of what underlying mechanisms may govern such pseudo-bifurcations. Hierarchical structure has been shown to dynamically influence system behavior \cite{Lera2020,OBrien2021,Wang2024}, motivating us to 
treat the non-normality parameter $\kappa$ as a potential control parameter for these phenomena. 
To explore this idea, we conducted simulations (see Methods) of: (i) one-dimensional normal forms approaching criticality via time-dependent control parameters, and (ii) a two-dimensional non-normal system with a constant spectrum but time-varying $\kappa$. Typical realisations are shown in the five top panels of Figure \ref{fig:early_warning}.

The bottom two panels of Figure \ref{fig:early_warning} show the time evolution of the effective eigenvalue and the variance for each simulation, calculated using a rolling window of 30 time units. Remarkably, these two indicators exhibit strikingly similar trends whether the underlying system is genuinely approaching a bifurcation in a normal system or experiencing an increase in non-normality, such as through a growing condition number. In both cases, the effective eigenvalue tends toward zero and the variance increases, mimicking classical early-warning signals. This convergence in behavior underscores a critical challenge: the same early-warning patterns can emerge from fundamentally different mechanisms, true critical slowing down near bifurcations or transient amplification driven by non-normal interactions, complicating the interpretation of such indicators in real-world systems.

The results confirm that systems with fixed spectra but increasing non-normality reproduce hallmarks of approaching bifurcations (\propref{pr:es_snl_not_fixed}), demonstrating that early-warning signals are not exclusive to bifurcations. Crucially, pseudo-bifurcations emerge at fixed $\kappa > \kappa_c$
in stationary systems \eqref{eq:basic_dynamics} fluctuating around stable fixed points, providing a minimal equilibrium model without requiring approach to criticality.
Given the prevalence of non-normal interaction matrices $\A$
in empirical systems \cite{OBrien2021}, pseudo-bifurcations likely occur far more frequently than true bifurcations. Indeed, systems typically encounter pseudo-bifurcations before genuine ones (\propref{pr:pseudo_bifurcation}). This is particularly relevant as complex systems often self-organize into hierarchical states that naturally generate non-normality \cite{Lera2020,OBrien2021,Wang2024}. With variance scaling as $\delta_\text{eff} \sim \kappa^2$ (\propref{pr:variance_reaction}) and autocorrelation time as $\tau_0 \sim \kappa^2$ (\propref{pr:slowing_down}), increasing non-normality amplifies both fluctuations and memory, heightening the risk of misidentifying pseudo-bifurcations as critical transitions.

\section*{Non-Normal Transients during Epileptic Seizure}

To illustrate how pseudo-bifurcations can manifest themselves in practice, we apply our framework to real EEG recordings from epileptic patients.

\begin{figure*}[!htb]
    \centering
    \includegraphics[width=0.9\linewidth]{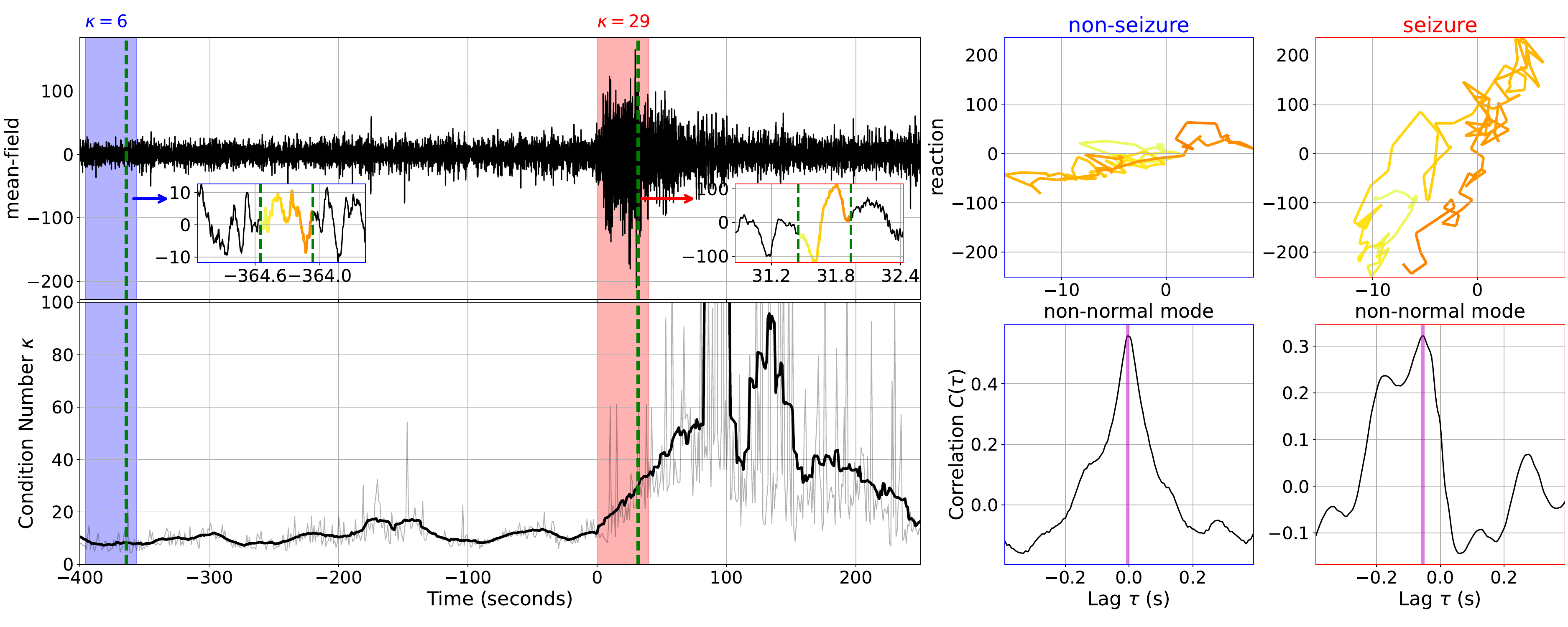}
    \caption{
        (Top Left)
        Mean-field state of the 23 EEG channels before, during (red), and after an epileptic seizure. 
        Inset plots show a high-resolution version of a single second before (blue) and during (red) the seizure.         
        (Bottom Left)
        Condition number $\kappa$ of the EEG connectome estimated over a rolling window of 40 seconds (light grey line).        
        An additional moving average over a time scale of 10 seconds is applied, resulting in the smoother black line.
         (Top Right) Loops, similar to Figure \ref{fig:simulations}(d), along the non-normal and reaction components during an excerpt of non-seizure (left/blue; stochastic loops) and seizure (right/red; quasi-deterministic loops).
        (Bottom Right)
        Lead-lag correlation between the system's exposure to the non-normal component and the mean-field state during a non-seizure (blue frame) and a seizure (red frame) period. 
        The purple vertical line indicates when the correlation reaches its maximum. 
        During the seizure, activity along the non-normal component consistently precedes the mean-field excursions by approximately 0.1 seconds.
    }
    \label{fig:empirical}
\end{figure*}

Brain connectomes are comprehensive maps of structural and functional neural connections.
They have become essential tools for studying the brain's complex dynamics.
In the context of epilepsy, connectomes have been used to investigate the neural underpinnings of seizures,
which are characterised by sudden, excessive synchronisation of neural activity \cite{Bullmore2011}.
It has been hypothesised that seizures may be associated with the brain operating near a critical 
threshold where neural dynamics are poised between order and chaos \cite{Royer2022}.
However, whether the brain truly operates at criticality during epileptic seizures remains a subject of debate \cite{Wilkat2019,Maturana2020}.
A mean-field model of excitatory and inhibitory dynamics has been proposed to explain these phenomena \cite{diSanto2018},
showing that non-normal transients that arise from the interplay between excitation and inhibition
can generate large bursts of neural activity reminiscent of seizure dynamics.
Building on these insights, and in light of evidence that brain connectomes are highly non-normal \cite{OBrien2021},
we hypothesise that non-normal transient bursts may underlie the observed increase in excitability during seizures,
shifting the brain into a more excitable and uncontrolled regime,
thereby revealing the intrinsic non-normal characteristics of the connectome.

To test this hypothesis, we analyse an EEG recording of a patient during an epileptic seizure,
collected at the Children's Hospital Boston \cite{Goldberger2000}.
The data include measurements from 23 EEG channels sampled at 256 Hz,
covering the time before, during, and after a seizure event lasting 40 seconds (see top-left panel of Figure \ref{fig:empirical} for the mean-field dynamics).
We empirically estimate the system's non-normality $\kappa$,
along with the non-normal and reaction modes. For this, we 
introduce a robust empirical method to estimate non-normal and reaction modes directly from data. It is based on solving an optimization problem 
maximizing a quadratic form involving the estimated interaction matrix, which
identifies two orthogonal directions, one capturing the system's most amplified response and the other its most sensitive direction. 
This approach bypasses the numerical instability of traditional decompositions and allows reliable detection of non-normal amplification in noisy, high-dimensional brain data
(see Methods).

The bottom-left panel of Figure \ref{fig:empirical} reveals a pronounced increase in the system's non-normality that coincides precisely with the onset of the epileptic seizure and persists long after its clinical termination.
This provides quantitative evidence that the seizure-related dynamics extend beyond the observable clinical symptoms.
This conclusion is supported by the sustained high amplitude and variance in the EEG signals, which remain elevated after the seizure has clinically ended.
Consistent with the increased non-normality observed after seizure onset, we detect quasi-deterministic loops during, but not prior to, the seizure (see top-right panel of Figure \ref{fig:empirical}).

By analysing the lead-lag correlation between the projection of the system onto the non-normal mode $\hn$ and the mean-field state $\bar{x}$,
we find that the non-normal component leads the mean-field dynamics by approximately 0.1 seconds (Figure \ref{fig:empirical}, bottom-right).
This suggests that epileptic seizures may not only be driven by non-normal dynamics, but that their onset could potentially be anticipated by monitoring the system's projection along the non-normal direction.
This insight opens promising avenues for early detection and potential control of seizure events, and may also provide strategies for managing self-induced instabilities in complex systems more broadly \cite{Troude2024}.

\begin{figure}[!htb]
    \centering
    \includegraphics[width=0.9\linewidth]{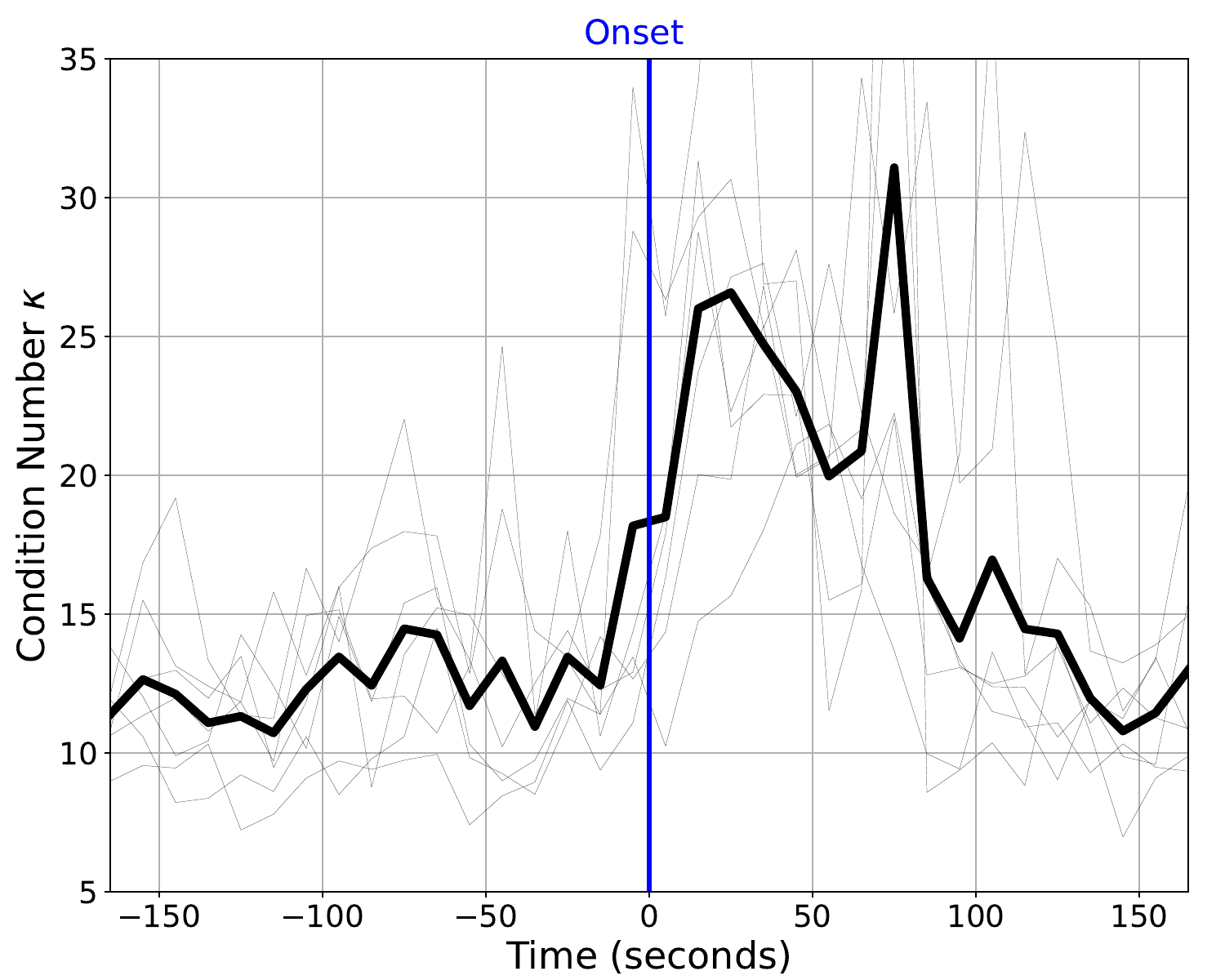}
    \caption{
        Condition number $\kappa$ of the EEG connectome around the onset of epileptic seizures,
        measured across six seizure events for patient chb01 in the dataset from \cite{Goldberger2000}.
        Thin grey lines represent individual seizures, while the thick black line shows the average.
        Time $t=0$ corresponds to the clinically defined seizure onset according to \cite{Goldberger2000}.
        }
    \label{fig:eeg_multiple}
\end{figure}

We also compute the condition number $\kappa$ as a function of time around the seizure onset for the same patient (chb01), across all six seizure events reported in the dataset \cite{Goldberger2000}. Each seizure is shown as a thin grey line in Figure \ref{fig:eeg_multiple}, with their average overlaid as a thick black line.
On average, the condition number increases during the epileptic state compared to the baseline (pre-seizure) period.
This suggests that, while epileptic seizures are known to produce early-warning signals, these signals may originate from the increasing non-normality of the
system (as illustrated in the numerical example of figure \ref{fig:early_warning}), rather than from proximity to a true bifurcation.

\section*{Discussion}

Near equilibrium, complex system dynamics are well approximated by equation \eqref{eq:basic_dynamics}, where the Jacobian matrix $\mathbf{A}$
is typically non-normal, being the rule rather than the exception. Any system approaching criticality first exhibits pseudo-bifurcations, making critical behavior a subset of this broader phenomenon. From a measure-theoretic perspective, normal operators have measure zero while non-normal operators are generic. 

While bifurcation theory provides crucial insights into critical transitions, it addresses only specific scenarios. Non-normal transients occur under much broader conditions, even far from bifurcation points. For system \eqref{eq:MF_AR1}, criticality requires the largest eigenvalue's real part near zero, whereas non-normal transients appear well below this threshold. Many phenomena interpreted as criticality may actually be non-normal transients.

True bifurcations do occur in systems like coral reefs transitioning to algae-dominated states under stress \cite{Mumby2007}, the collapsed Newfoundland cod fishery \cite{Hutchings1994}, and potentially the AMOC \cite{Boulton2014}. However, pseudo-bifurcations are equally plausible in climate systems, given established non-normal behavior in hydrodynamic stability \cite{ButlerFarrell1992,FarrellIoannou1993,ConstantinouFarrellIoannou2014,FarrellIoannou2014}. Non-normal approaches have proven valuable for climate predictability, including stochastic optimal forcing of ENSO \cite{Flugel2004}, with broader applications in fluid dynamics reviewed in \cite{Schmid2007}.

The assumption of criticality often remains hypothetical. Epileptic seizures exemplify this debate \cite{Wilkat2019,Maturana2020}, with our findings supporting the non-critical perspective. Moreover, non-normality can accelerate escape from equilibria even far from bifurcations, enabling rapid transitions between metastable states \cite{troude2025}.

In summary, we have expanded early-warning signal applications beyond bifurcations. High non-normality systems exhibit pseudo-bifurcations with signatures identical to true bifurcations (critical slowing down, increased variance) without indicating proximity to criticality. Our EEG analysis during epileptic seizures demonstrates non-normal dynamics' dominant role. Many systems thought near criticality may actually remain near equilibrium, with extremes arising from non-normal transients rather than bifurcations, necessitating reexamination of supposedly critical systems.

The following table provides a synopsis of our main results.
\begin{center}
\fbox{%
  \begin{minipage}{0.45\textwidth}
    \footnotesize
    \textbf{Key Results}
    \begin{itemize}
      \item \textbf{Non-normal dimension reduction}:  
        All transient amplification in an \(N\)-dimensional stochastic non-normal linear system (\defref{def:snl}) can be captured by projecting onto a two-dimensional subspace spanned by the non-normal mode \(\hn\) and its reaction \(\hr\) (\propref{pr:dim_red}).
      \item \textbf{Towards the unit root}:  
        In the reduced OU form \( \dot r_t = -\theta_t\,r_t + \sqrt{2\delta}\,\eta_t\), the time-dependent coefficient \(\theta_t\) can transiently 
        become negative when \(\kappa>\kappa_c\). This mimics the behavior in discrete time of a system approaching a unit root and gives rise to pseudo-bifurcations (\propref{pr:effect_ou}).
      \item \textbf{Variance growth}:  
        The variance along the reaction scales as \(\mw{r^2}\sim\kappa^2\) for large condition number \(\kappa\),
        reproducing the ``increasing variance'' early-warning signal (\propref{pr:variance_reaction}) in systems in which $\kappa$ 
        increases.
      \item \textbf{Slowing down}:  
        The instantaneous autocorrelation time scales as \(\tau_0\sim\kappa^{2}\) for large condition number \(\kappa\),
        resulting in prolonged memory and ``slowing down'' of the dynamics (\propref{pr:slowing_down})  in systems in which $\kappa$ 
        increases.
        \item  \textbf{Onset of non-normal early-warning signals}:
The critical condition number $\kappa_c$ approaches $1$ as the system nears bifurcation, implying that non-normal early-warning signals, such as increased variance, autocorrelation, and reactivity, emerge and intensify well before the system exhibits spectral signatures of a true bifurcation.
    \end{itemize}
  \end{minipage}%
}
\end{center}

Stanislaw Ulam famously quipped that ``using a term like nonlinear science is like referring to the bulk of zoology as the study of non-elephant animals,'' humorously highlighting that linear systems are rare in nature while nonlinear dynamics dominate real-world phenomena. This remark underscores how the emphasis on linearity has long overshadowed the vastly more prevalent and complex behaviors of nonlinear systems. We propose that a similar oversight has occurred with non-normal dynamics. In complex systems, interaction matrices are typically non-normal, making transient dynamics induced by non-normality not rare anomalies but fundamental features of natural systems. These transient behaviors exhibit remarkable universality, emerging even in linear non-normal systems. Just as the field has evolved from its historical overemphasis on linearity to embrace nonlinear dynamics, we suggest it is time to move beyond the narrow focus on bifurcations to recognize the broader and more pervasive role of non-normal transients in shaping complex system behavior.

\section*{Materials and Methods}

The dimensional reduction leading to the set of two equations \eqref{eq:dim_red} is proved in the \textit{Supplementary Material}.
All subsequent results showing that non-normal systems reproduce early-warning signals (\defref{def:early_warning})
are also proved in the \textit{Supplementary Material}.

In this section, we present the methods behind both the numerical and empirical analyses.

\subsection*{Non-Normal System Mimicking Pre-Bifurcation Behaviours}

To generate Figure \ref{fig:early_warning}, we analyzed four canonical one-dimensional bifurcation normal forms augmented with additive noise
\begin{subequations}    \label{eq:normal_forms}
    \begin{align}
        &\dot{x}_t = x_t^2 - \frac{1}{4}\lambda_{\text{eff}}^2 +\sqrt{2\delta}~\eta_t
        &&\textbf{(Fold)} \\
        &\dot{x}_t = \lambda_{\text{eff}}x_t - x_t^2 +\sqrt{2\delta}~\eta_t
        &&\textbf{(Transcritical)} \\
        &\dot{x}_t = \lambda_{\text{eff}}x_t - x_t^3 +\sqrt{2\delta}~\eta_t
        &&\textbf{(Pitchfork, Supercritical)} \\
        &\dot{x}_t = \lambda_{\text{eff}}x_t + x_t^3 +\sqrt{2\delta}~\eta_t
        &&\textbf{(Pitchfork, Subcritical)}
    \end{align}
\end{subequations}
where $\eta_t \overset{\textit{i.i.d}}{\sim} \mathcal{N}(0,1)$.
The four bifurcations considered are fold, transcritical, and both supercritical and subcritical pitchforks.

Here, $\lambda_{\text{eff}}$ acts as the control parameter. In the Ornstein-Uhlenbeck (OU) approximation:
\begin{equation}
    \dot{x}_t = \lambda_{\text{eff}}(x_t-x^*) + \sqrt{2\delta}~\eta_t + \mathcal{O}((x_t - x^*)^2),
\end{equation}
where $x^*$ denotes the stable fixed point, the term $\lambda_{\text{eff}}$ corresponds to the mean-reversion rate. Hence, in the OU approximation, the variance is given by:
\begin{equation}
    \mw{x^2} = \frac{\delta}{|\lambda_{\text{eff}}|}.
\end{equation}

We consider a time-dependent control parameter of the form:
\begin{equation}
    \lambda_{\text{eff},t} = vt - \lambda_0 < 0,
\end{equation}
so that the system approaches the bifurcation point linearly in time.

Next, we consider a non-normal system governed by equation \eqref{eq:dim_red}, for which the variance along the reaction coordinate is given by expression \eqref{eq:variance_reaction}.
We seek a time-dependent condition number $\kappa_t$ such that the theoretical variance of the reaction matches, at all times, the variance of the OU process:
\begin{equation}    \label{eq:var_k_lambda}
    \mw{r} = \frac{\delta\alpha}{\alpha^2 - 1}\cdot \frac{\kappa_t^2 + 1}{2}
    \left[1 - \frac{\kappa_t^2 - 1}{\kappa_t^2 + 1} \cdot \frac{\alpha^2 - 1}{\alpha} \right] 
    = \frac{\delta}{\lambda_{\text{eff},t}}.
\end{equation}

At time $t = 0$, we set the system to be normal, i.e. $\kappa_0 = 1$, and choose $\alpha$ such that:
\begin{equation}
    \alpha = \frac{|\lambda_0|}{2} + \sqrt{\left(\frac{\lambda_0}{2}\right)^2 + 1}.
\end{equation}
Recalling that $\lambda_{\text{eff},t} = vt - \lambda_0$, so that $\lambda_{\text{eff},0} = -\lambda_0$,
we then determine $\kappa_t$ by solving \eqref{eq:var_k_lambda} at each time point:
\begin{equation}
    \kappa_t = \sqrt{ \frac{\alpha}{\alpha - \lambda_0} \left[ 2 \cdot \frac{\lambda_0}{|\lambda_{\text{eff},t}|} - \frac{\lambda_0}{\alpha} - 1 \right] }.
\end{equation}

This procedure ensures that all normal forms in \eqref{eq:normal_forms} and the reaction coordinate of the non-normal dynamics in \eqref{eq:dim_red}
share the same theoretical variance across time.

Figure \ref{fig:early_warning} presents simulations of all systems under the same time-dependent control parameters,
showing that the non-normal system -- with a fixed spectral structure -- perfectly mimics the early-warning signal dynamics exhibited by normal forms as they approach bifurcation.

\subsection*{Empirical Estimation of Non-Normal and Reaction Modes}

For our analysis of the brain connectome, 
we estimate the matrix $\A$ in equation \eqref{eq:basic_dynamics} as $\hat{\A}$ obtained by fitting a VAR(1) process to the EEG data over rolling windows of 40 seconds. 
In principle, we could then proceed by first determining the transition matrix $\PP$ that diagonalizes $\hat{\A}$, 
and subsequently calculating the SVD $\PP=\mathbf{U} \Sigma \mathbf{V}^\dagger$ to identify the non-normal mode $\hn \sim \hu_N$ and reaction mode $\hr \sim \sum_{i=1}^{N-1} \hu_i$. 
However, as we explain in detail in the \textit{Supplementary Material}, this approach is highly sensitive to noise, especially when $\A$ is non-normal. 
Therefore, we propose an alternative approach that consists in estimating the non-normal and reaction component directly from $\hat{\A}$ as
\begin{equation} \label{eq:n_e_optimization}
    (\hr, \hn)
    ~=~
    \argmax_{ \mathbf{v}, \mathbf{w} \in \mathbb{S}^{N-1} }~ \langle \mathbf{v}, \hat{\A} \mathbf{w} \rangle  
    ~~ \text{s.t.}~~
    \langle \mathbf{v}, \mathbf{w} \rangle = 0. 
\end{equation}
where $\mathbb{S}^{N-1}$ is the unit sphere in $N$ dimensions.
Here, 
$\langle \mathbf{v}, \hat{\A} \mathbf{w} \rangle$ represents the inner product between the vector $\mathbf{v}$ and the result of applying the matrix $\hat{\A}$ to the vector $\mathbf{w}$. 
It thus measures the degree of alignment or projection of the transformed vector $\hat{\A} \mathbf{w}$ onto $\mathbf{v}$.
Equation \eqref{eq:n_e_optimization} directly captures the intuition behind the reduced equation \eqref{eq:dim_red} whereby the non-normal component maximally projects onto the reaction component.
From \eqref{eq:dim_red}, we further see that $\kappa$ can be estimated directly from $\hr$ and $\hn$ via 
$\kappa 
= 
\sqrt{
\left| \langle \hr, \hat{\A} \hn \rangle \right| / \left|  \langle \hn, \hat{\A} \hr \rangle \right|
}
$ 
thus allowing us to completely bypass the need for estimating the singular values which are numerically unstable.
In the \textit{Supplementary Material}, we test our empirical procedure on synthetic data, finding that our procedure faithfully recovers the non-normal and reaction mode, 
and estimates $\kappa$  more precisely than when calculating it directly via singular value decomposition.

\balance
\bibliographystyle{naturemag}  
\bibliography{bibliography} 

\section*{Acknowledgements}

This work is partially supported by the National Natural Science Foundation of China (Grant No. T2350710802 and No. 72350410487), 
Shenzhen Science and Technology Innovation Commission Project (Grant No. GJHZ202107-05141805017 and No. K23405006), 
and the Center for Computational Science and Engineering at Southern University of Science and Technology.



\clearpage
\onecolumn
\appendix

\section*{\Large\bf Supplementary Materials}
\setcounter{section}{0}
\renewcommand{\thesection}{S\arabic{section}}
\renewcommand{\thesubsection}{S\arabic{section}.\arabic{subsection}}

\tableofcontents

\newpage


\section{Definition of $n$-dimensional linear stochastic system}

Throughout this Supplementary Material, we consider an $n$-dimensional linear stochastic system of the form
\begin{equation}    \label{eq:system}
    \dot{\x}_t ~=~ \A~\x_t + \sqrt{2\delta}~ \et_t,
    \quad \et_t~\overset{\textit{i.i.d}}{\sim}~\mathcal{N}(0,1)
\end{equation}
The $N \times N$  matrix $\A$, which governs the interactions of the system,
is assumed diagonalizable with eigenvalues $\lambda_1, \ldots, \lambda_N$ all having negative real parts, 
thus ensuring that the equilibrium point $\x=0$ remains stable.
The diagonal $N$-dimensional noise term $\et_t$ is assumed to be normally distributed, and the scalar $\delta > 0$ is a scaling factor.

\section{Proof of Non-Normal Dimensional Reduction}
\label{SI:dimension_reduction}

In this section, we systematically derive the representation of the non-normal system \eqref{eq:system} as an effective two-dimensional dynamics along the non-normal and reaction mode. 
We first provide a short review of the singular value decomposition that our analysis relies upon. 
We then provide some intuition for the emergence of transients, followed by a detailed derivation of the two effective system components. 

\subsection{Singular Value Decomposition}

Because $\A$ is diagonalizable, there exists a matrix $\PP$ which is the eigenbasis transformation of $\A$ such that
\begin{equation} \label{eq:A_diag}
    \A = \PP\Lam\PP^{-1}
\end{equation}
where $\Lam = \text{Diag}\left(\lambda_1,\cdots,\lambda_N\right)$ is the diagonal matrix composed of the eigenvalues of $\A$.
Here, we assume that $\lambda_i < 0~\forall i$ and we explicitly set the minus sign to indicate that the eigenvalues are negative. 

For $\A$ normal, that is, for $\A \A^\dagger = \A^\dagger \A$, then $\PP$ is unitary and $\A$ can be diagonalized by means of a rotation. 
But in general, we assume here that $\A$ is non-normal, and hence $\PP$ is not unitary. 
Quite generally, a matrix $\PP$ can be represented as a sequence of a rotation, a rescaling, and another rotation, via the singular value decomposition (SVD) given by  
\begin{equation}    \label{eq:SVD_P}
    \PP = \U\Sig\V^\dag
\end{equation}
where $\U$ and $\V$ are unitary matrices, $\V^\dagger$ is the Hermitian conjugate of $\V$ and $\Sig = \text{diag}(\sigma_1,\cdots,\sigma_N)$ is a diagonal matrix of the singular values $\{ \sigma_i \}$. 
Without loss of generality, we can assume that $\sigma_1 \geqslant \sigma_2 \geqslant \ldots \geqslant \sigma_N>0$ because a permutation of a unitary matrix remains unitary.  

Combining \eqref{eq:A_diag} with \eqref{eq:SVD_P} we can write 
\begin{equation} \label{eq:SVD_A}
    \A
    = \U\Sig\V^\dag\Lam\V\Sig^{-1}\U^\dag 
    = \sum_{i,j=1}^n \frac{\sigma_i}{\sigma_j}\lambda_{ij} \hu_i\hu_j^\dag
    ,\qquad \text{with }
    \lambda_{ij} = -\sum_{k=1}^n \lambda_k v_{ik} v_{jk}^*
    ,
\end{equation}
and $\hu_i$ and $\hv_i$ denote the $i^{\text{th}}$ column of the unitary matrix $\U$ and $\V$, respectively.
Further, $v_{ij}$ is the element of $\V$ at the $i^{\text{th}}$ column and $j^{\text{th}}$ row and $v_{ij}^*$ is the complex conjugate of $v_{ij}$. 

Expression \eqref{eq:SVD_A} is crucial for our subsequent derivations, since it shows how the matrix $\A$ can be decomposed into $N^2$ rank-1 projections $\hu_i \hu_j^\dagger$ modulated by a factor $\frac{\sigma_i}{\sigma_j} \lambda_{ij}$ . 
As we will see below, the factor $\lambda_{ij}$ is of order one, such that the dominant contributions to $\A$ will come from components where $\sigma_i / \sigma_j$ is large. 

\subsection{Effective Two-Dimensional Dynamics}

We now show how a system \eqref{eq:system} that is non-normal along one dominant component can be decomposed into an effective two-dimensional dynamics. 
To this end, recall that the SVD \eqref{eq:SVD_P} of $\PP$ is composed of two rotations with a scaling along the singular values $\Sig$ in-between.
The greater the deviation of these singular values from $1$, the more pronounced the scaling effect. For the remainder of this article, we assume that only the final dimension has a singular value significantly smaller than $1$, while all other dimensions are roughly equal to one, that is
$1 = \sigma_1 \gtrsim \sigma_2 \gtrsim \ldots \gtrsim \sigma_{N-1} \gg \sigma_N$. 
This assumption essentially encapsulates the non-normal behavior along one component. 
We further define matrix $\A$'s condition number as $\kappa = \sigma_1 / \sigma_N \gg 1$.
For the remaining singular values, $i=2, \ldots, N-1$, we write $\sigma_i = (1-\delta_i) \sigma_1$ with $\delta_i \ll 1$ such that $\sigma_i$ is slightly less than but close to $\sigma_1=1$. 
To leading order, we can thus write 
\begin{equation}
    \frac{\sigma_i}{\sigma_j} = 
    \begin{cases}
        1 + (\delta_{j} - \delta_i) + \mathcal{O}(\delta_i\delta_j, \delta_j^2)
        & \text{if } i,j \neq N \\
        \kappa(1 - \delta_i)
        & \text{if } i \neq N, j = N \\
        \kappa^{-1} + \mathcal{O}(\kappa^{-1}\delta_j)
        & \text{if } i = N, j \neq N \\
        1
        & \text{if } i = j = N. 
    \end{cases}
\end{equation}
Defining further $\delta^2 = \sum_{i}\delta_i ^2$ and plugging these approximations into \eqref{eq:SVD_A} allows us to decompose the matrix $\A$ as 
\begin{equation} \label{eq:A_decomposition}
    \A 
    =
    \A'
    + 
    (\kappa-1)a_{nr}\hr\hn^\dag
    + \kappa\delta \mathbf{e}\hn^\dag 
    + (\kappa^{-1}-1)a_{nr}\hn\hr^\dag
    + \delta\mathbf{E}
    + \mathcal{O}(\delta^2, \kappa^{-1}\delta),
\end{equation}
where 
\begin{equation}
    \A' = \U\V^\dag \Lam\V\U^\dag,
\end{equation}
is the ``normal component'' of $\A$. 
Moreoever, we have defined the non-normal mode $\hn = \hu_N$, and the reaction mode $\hr$ via 
\begin{equation}
    a_{nr} \hr = \sum_{i=1}^{N-1}\lambda_{iN}\hu_i
    \quad\text{where}\quad
    \|\hr\| = 1. 
\end{equation}
The matrix $\mathbf{E}$ and vector $\mathbf{e}$ are given by 
\begin{equation}
   \mathbf{E} = {1 \over  \delta} \sum_{i,j=1}^{N-1}(\delta_j - \delta_i)\lambda_{ij}\hu_i\hu_i ^\dag, 
    \quad
    \mathbf{e} = {1 \over  \delta}  \sum_{i=1}^{N-1}\delta_i\lambda_{iN}\hu_i,
\end{equation}
and encapsulate the comparatively small interactions resulting from the anisotropy of the small deviation $\delta_i$ from non-normality along the $N-1$ dimensions.
We can see from equation \eqref{eq:A_decomposition}
that $\A$ is decomposed into a normal part as well as off-diagonal parts which are dominated by a term of order $\kappa$. 
Since these dominating interactions arise along $\hr$ and $\hn$, we can approximately restrict our study to these two dimensions,
and project the matrix $\A$ into $(\hn,\hr)$ to define
\begin{equation}  \label{eq:reduce_A}
    \Gam' \equiv 
    \begin{pmatrix}
        a_n & \kappa^{-1} a_{nr} \\
        \kappa a_{rn} & a_r
    \end{pmatrix}
    + \mathcal{O}(\delta^2, \kappa^{-1}\delta) 
\end{equation}
with the matrix coefficients of $\Gam'$ given by
\begin{subequations} \label{eq:reduce_A_para}
    \begin{align}
        & a_n = \lambda_{NN} 
        && a_{nr} = \sqrt{\sum_{i=1}^{N-1} |\lambda_{iN}|^2}
        \\
        & a_r = a_r ^0 + \delta a_r ^1
        && a_{rn} = a_{nr} + \delta a_{rn}^1 \\
        & a_r ^0 = \frac{1}{|a_{nr}|^2}\sum_{i,j=1}^{N-1}\lambda_{ij}\lambda_{iN}^*\lambda_{jN}
        && a_r ^1 = \frac{1}{|a_{nr}|^2}\sum_{i,j=1}^{N-1}\frac{\delta_j - \delta_j}{\delta}\lambda_{ij}\lambda_{iN}^*\lambda_{jN} \\
        &a_{rn}^1 = \frac{1}{a_{nr}}\sum_{i=1}^{N-1}\frac{\delta_i}{\delta}|\lambda_{iN}|^2
        && \lambda_{ij} = -\sum_{k=1}^N \lambda_k v_{ik}v_{jk} ^*.
    \end{align}
\end{subequations}
To summarize, we have shown that system \eqref{eq:system} can be reduced to the two-dimensional system 
\begin{equation}    \label{eq:systemqregqr}
   \frac{d}{dt}
    \begin{pmatrix}
    n_t  \\
    r_t
    \end{pmatrix}
    ~=~  \Gam'~
    \begin{pmatrix}
    n_t  \\
    r_t
    \end{pmatrix} ~+~ \sqrt{2\delta} ~ 
    \begin{pmatrix}
    \eta_{n,t}  \\
    \eta_{r,t}
    \end{pmatrix}
   ~~,   ~~~{\rm with}~  \Gam ~{\rm defined~by}~\eqref{eq:reduce_A}  
\end{equation}
where the non-normal ($n$) and reaction ($r$) components are respectively the projection along $\hn$ and $\hr$.

\subsection{Reduced $2\times 2$ Non-Normal Form}

We have demonstrated that any $N$-dimensional model can be reduced to two effective dimensions
whose dynamics is governed by the reduced matrix $\Gam'$ \eqref{eq:reduce_A}, which
captures the essential non-normal behaviour \eqref{eq:dim_red}.  
Assume $\Gam'$ is diagonalizable with right-eigenvectors $\p_\pm$, so that the eigenbasis matrix is
$\PP=(\p_+,\,\p_-)$ and  
\begin{equation}
    \Gam'=\PP\Lam\PP^{-1}
    ,\qquad
    \Lam=\operatorname{diag}(\lambda_+,\lambda_-).
\end{equation}

As for the general $N$-dimensional case, write the SVD of $\PP$ as $\PP=\U\Sig\V^\dag$,
where $\U$ and $\V$ are unitary and we can write the diagonal matrix $\Sig$ composed of the singular values of $\PP$ as
\begin{equation}
    \Sig = \sqrt{\sigma_+\sigma_-}\;
            \K^{1/2},
    \qquad
    \K=\begin{pmatrix}\kappa&0\\[2pt]0&\kappa^{-1}\end{pmatrix},
    \qquad
    \kappa=\frac{\sigma_+}{\sigma_-}\ge 1,
\end{equation}
by convention. Because the eigenvectors are normalized i.e. $\|\p_\pm\|=1$,  
\begin{equation}
    \PP^\dag\PP ~=~ \V\Sig^2\V^\dag ~=~ \begin{pmatrix} 1 & c \\ c^* & 1 \end{pmatrix},
    \qquad
    c ~=~ \mw{\p_+,\p_-},
\end{equation}
so that
\begin{equation}
    \sigma_\pm=\sqrt{1\pm|c|},\qquad
    \kappa=\sqrt{\frac{1+|c|}{1-|c|}},\qquad
    \V=\frac{1}{\sqrt{2}}
        \begin{pmatrix}e^{i\phi}&-e^{i\phi}\\ 1&1\end{pmatrix},
    \quad
    \phi=\arg(c) .
\end{equation}

Insert the decomposition into $\Gam'$:
\begin{align}
    \Gam'
   &=\PP\Lam\PP^{-1}
     =\U\Sig\V^\dag\Lam\V\Sig^{-1}\U^\dag\notag\\
   &=\frac{1}{2}\U\K^{1/2}
     \begin{pmatrix}
        1 & e^{-i\phi}\\[2pt] 1 & -e^{-i\phi}
     \end{pmatrix}
     \Lam
     \begin{pmatrix}
        e^{i\phi} & -e^{i\phi}\\[2pt] 1 & 1
     \end{pmatrix}
     \K^{-1/2}\U^\dag
     =\U\Gam_u\U^\dag ,
\end{align}
with
\begin{equation}
    \Gam_u=
    \begin{pmatrix}
        \alpha_+ & \alpha_-\kappa\\[2pt]
        \alpha_-\kappa^{-1} & \alpha_+
    \end{pmatrix},
    \qquad
    \alpha_\pm=\frac{1}{2}(\lambda_+\pm\lambda_-).
\end{equation}

Provided the spectrum is real, choose the ordering $0>\lambda_+>\lambda_-$ so that $\alpha_->0$.  
Rescaling by $\alpha_-$ and applying $\U^\dagger$ yields the universal form
\begin{equation}
    \Gam =\frac{1}{\alpha_-}\U^\dagger\Gamma\U
    =
    \begin{pmatrix}
        -\alpha & \kappa\\ \kappa^{-1} & -\alpha
    \end{pmatrix},
    \qquad
    \alpha=-\frac{\alpha_+}{\alpha_-}.
\end{equation}
Hence only two independent parameters remain:
\begin{itemize}
    \item $\alpha$ measures the spectral distance from criticality
    ($\lambda_\pm/\alpha_-=-\alpha\pm1$, so criticality corresponds to $\alpha=1$);
    \item $\kappa\ge1$ quantifies the degree of non-normality.
\end{itemize}

Using the time scaling $t\mapsto\alpha_-t$ and the unitary transformation
$\U$, the reduced stochastic system becomes
\begin{subequations}    \label{eq:apx_nnf}
    \begin{align}
        \dot{n}_t &~=~ -\alpha n_t + \kappa^{-1} r_t + \sqrt{2\delta}~\eta_{n,t},\\
        \dot{r}_t &~=~ -\alpha r_t + \kappa      n_t + \sqrt{2\delta}~\eta_{r,t},
    \end{align}
\end{subequations}
where $\eta_{n,t},\eta_{r,t}$ are standard independent white noises.

Because $r_t$ dominates the transient amplification when $\kappa\gg 1$, one may integrate \eqref{eq:apx_nnf} to obtain
\begin{equation}    \label{eq:apx_reaction_red}
    r_t ~=~ \sqrt{2\delta}\int_{0}^{t}e^{-\alpha(t-s)}
         \bigl[\kappa\sinh(t-s)\,\eta_{n,s}
              +\cosh(t-s)\,\eta_{r,s}\bigr]~ds.
\end{equation}
Equivalently, introducing a single \textit{i.i.d.} noise $\eta_t\sim\mathcal N(0,1)$ gives
\begin{equation}\label{eq:apx_reaction_rnnf}
    r_t=\sqrt{2\delta}\int_{0}^{t}e^{-\alpha(t-s)}
         \sqrt{\kappa^{2}\sinh^{2}(t-s)+\cosh^{2}(t-s)}\;\eta_s\,ds,
\end{equation}
which is the compact reduced description governing the leading-order non-normal dynamics.

\subsection{Summary of results}

We showed that any $N$-dimensional stochastic non-normal linear system \eqref{eq:system} with a diagonalizable interaction matrix $\A$ can be reduced to an effective $2$–dimensional model when the following spectral structure holds:
\begin{itemize}
	\item the eigenbasis matrix $\PP$ has $N-1$ singular values that are nearly identical,
	$|\sigma_i-\sigma_j|\ll 1$ for $i,j=1,\ldots, N-1$;
	\item the remaining singular value is much smaller,
	$\sigma_i\gg\sigma_N>0$ for $i=1,\dots,N-1$.
\end{itemize}

Under these conditions, the full dynamics is captured by the $2\times 2$ system
\begin{subequations}
\begin{align}
\dot{n}_t &~=~ -\alpha, n_t + \kappa^{-1} r_t + \sqrt{2\delta}~\eta_{n,t},\\
\dot{r}_t &~=~ -\alpha r_t + \kappa      n_t + \sqrt{2\delta}~\eta_{r,t},
\end{align}
\end{subequations}
where $\eta_{n,t}$ and $\eta_{r,t}$ are independent standard white noises.

The reaction coordinate $r_t$, which carries the leading $\mathcal O(\kappa)$ amplification,
can be written as
\begin{equation}
r_t ~=~ \sqrt{2\delta}~\int_{0}^{t} e^{-\alpha(t-s)}
\bigl[\kappa\sinh(t-s)\eta_{n,s} + \cosh(t-s)\eta_{r,s}\bigr]~ds .
\end{equation}
Introducing a single \textit{i.i.d.} noise $\eta_s\sim\mathcal{N}(0,1)$ gives the compact,
equivalent representation
\begin{equation}
r_t ~=~ \sqrt{2\delta}~\int_{0}^{t} e^{-\alpha(t-s)}
\sqrt{\kappa^{2}\sinh^{2}(t-s)+\cosh^{2}(t-s)};\eta_s~ds ,
\end{equation}
which fully characterises the dominant non-normal dynamics.


\section{Early-Warning Signals in Stochastic Non-Normal Linear Systems}


In a deterministic dynamical system $\dot{x} = f(x;\mu)$,
a bifurcation occurs when a smooth variation of a control parameter $\mu$ drives an equilibrium $x^*$ from stable to marginal stability
-- that is, when an eigenvalue of the Jacobian $J(\mu)=\partial_x f|_{x=x^{*}}(\mu)$ crosses the imaginary axis.
Mathematically, if $\lambda(\mu)$ denotes the right–most eigenvalue of $J(\mu)$,
the bifurcation point $\mu_c$ is defined by $\operatorname{Re}\lambda(\mu_c)=0$.
In stochastic or noisy data, one cannot observe the spectrum directly,
so one monitors three early-warning indicators that follow from the linear Ornstein–Uhlenbeck approximation
\begin{equation}
\dot{x}_t = \lambda(\mu)\,\bigl(x_t-x^{\!*}\bigr)+\sqrt{2\delta}\,\eta_t,\qquad \eta_t\sim\mathcal N(0,1), 
\qquad \lambda(\mu) \le 0.
\end{equation}

\begin{itemize}
	\item[E1] \textbf{Drift toward a unit root}:
	As $\mu\to\mu_c$, $\lambda(\mu)\to 0^{-}$;
	equivalently the discrete-time autoregressive coefficient
	$\phi = e^{\lambda(\mu)\Delta t}$ approaches the ``unit root'' $1$,
	signalling loss of mean-reversion.
	\item[E2] \textbf{Diverging variance}:
	The stationary variance satisfies $\mw{x^2}=\delta/|\lambda(\mu)|$,
	so $\mw{x}\to\infty$ as $\lambda(\mu)\to 0^{-}$.
	\item[E3] \textbf{Critical slowing down}:
	The lag-$\tau$ autocorrelation is $C(\tau)=e^{\lambda(\mu)\tau}$;
	its characteristic decay time $\tau_c=1/|\lambda(\mu)|$ diverges,
	implying progressively slower recovery from perturbations.
\end{itemize}
Detecting these three trends $\lambda \to 0^{-}$, $\mw{x^2}\uparrow$, and $\tau_c \uparrow$ 
in empirical data in real-time provides quantitative evidence that the system is approaching a bifurcation.

This section shows that a non-normal system,
when examined along its reaction coordinate,
displays exactly the same signatures as a system approaching a bifurcation,
even though its spectrum remains fixed.
These effects arise solely from the system's non-normality and are demonstrated using the reduced two-dimensional form  \eqref{eq:apx_nnf}.

\subsection{Proof: Drift Toward a Unit Root}

Starting from the reduced reaction dynamics \eqref{eq:apx_reaction_rnnf},
we rewrite $r_t$ compactly as
\begin{equation} \label{eq:Omega_def}
  r_t ~=~ \sqrt{2\delta}~\int_{0}^{t} e^{-\Omega(t-s)}\eta_s~ds,
  \qquad 
  \Omega(t)~=~
  \alpha t
  -\frac{1}{2}\ln \Bigl[\kappa^{2}\sinh(t)^{2}+\cosh(t)^{2}\Bigr],
\end{equation}
so that $r_t$ is an Ornstein-Uhlenbeck process whose mean-reversion rate $\theta(t) :=\!\Omega'(t)$ is itself time-dependent.  
The system therefore imitates critical behaviour whenever $\theta(t)<0$ for some $t$.

The minimum of $\Omega(t)$ occurs at the root of $  \theta(t)$ where
\begin{equation}\label{eq:theta_root2}
  \theta(t) ~=~
  \alpha
  -\frac{(\kappa^{2}+1)\cosh t~\sinh t}{\kappa^{2}\sinh(t)^{2}+\cosh(t)^{2}}~.
\end{equation}
By writing \eqref{eq:theta_root2} as a quadratic polynomial in
$e^{2t}$, the equation  $\theta(t) =0$ leads to
\begin{equation}
  (\alpha-1)\frac{\kappa^{2}+1}{4}~e^{-2t}
  \Bigl[
      e^{4t}
      -2~\frac{\kappa^{2}-1}{\kappa^{2}+1}~
        \frac{\alpha}{\alpha-1}~
        e^{2t}
      +\frac{\alpha+1}{\alpha-1}
  \Bigr] ~=~ 0,
\end{equation}
which yields the admissible (real and positive) solution
\begin{equation}\label{eq:e2t_star}
  e^{2t^\ast} ~=~
  \frac{\displaystyle \alpha}{\displaystyle \alpha-1}~
  \frac{\displaystyle \kappa^{2}-1}{\displaystyle \kappa^{2}+1}
  +\sqrt{
        \Bigl(\frac{\alpha}{\alpha-1}~
               \frac{\kappa^{2}-1}{\kappa^{2}+1}\Bigr)^{2}
        -\frac{\alpha+1}{\alpha-1}}~.
\end{equation}
A real root exists only when the discriminant is non-negative,
which requires
\begin{equation}\label{eq:kappa_c_exact}
  \kappa ~\ge~ \kappa_{c}
  ~:=~
  \alpha+\sqrt{\alpha^{2}-1}
  ~=~
  \begin{cases}
    2(\alpha-1)+\mathcal O\bigl((\alpha-1)^{-1}\bigr), & \alpha\gg 1,\\
    1+\sqrt{2(\alpha-1)}+\mathcal O(\alpha-1),          & \alpha\gtrsim 1 .
  \end{cases}
\end{equation}

Far from criticality ($\alpha\gg 1$),
when $\kappa\gg 2\alpha$, expanding \eqref{eq:e2t_star} gives
\begin{equation}
    t^\ast
    ~=~
    \frac{1}{2}\ln\Bigl(\frac{\alpha+1}{\alpha-1}\Bigr)
    -\frac{\alpha}{\kappa^{2}}
    +\mathcal O(\kappa^{-4}),
\end{equation}
and the minimal mean-reversion rate becomes
\begin{equation}\label{eq:Omega_min}
  \Omega_{\min} ~:=~ \Omega(t=t^*)
  ~=~
  \frac{1}{2}\ln\!\Bigl[\frac{(\alpha+1)^{\alpha+1}}{(\alpha-1)^{\alpha-1}}\Bigr]
  -\ln\kappa
  +\mathcal O\!\bigl((\alpha/\kappa)^{2}\bigr).
\end{equation}
Thus $\Omega_{\min}<0$ (i.e.\ local loss of mean reversion) whenever
\begin{equation}
    \kappa ~\gtrsim~ \frac{(\alpha+1)^{(\alpha+1)/2}}{(\alpha-1)^{(\alpha-1)/2}} ~\approx~ e(\alpha-1)~,\qquad \alpha\gg1 .
\end{equation}
Near criticality ($\alpha\gtrsim 1$), the condition simplifies to
\begin{equation}
    \kappa ~\gtrsim~ 2
    +\bigl[1+\ln 2\bigr](\alpha-1)
    +\mathcal O\!\bigl((\alpha-1)\ln(\alpha-1)\bigr).
\end{equation}

\begin{figure}
    \centering
    \includegraphics[width=0.75\textwidth]{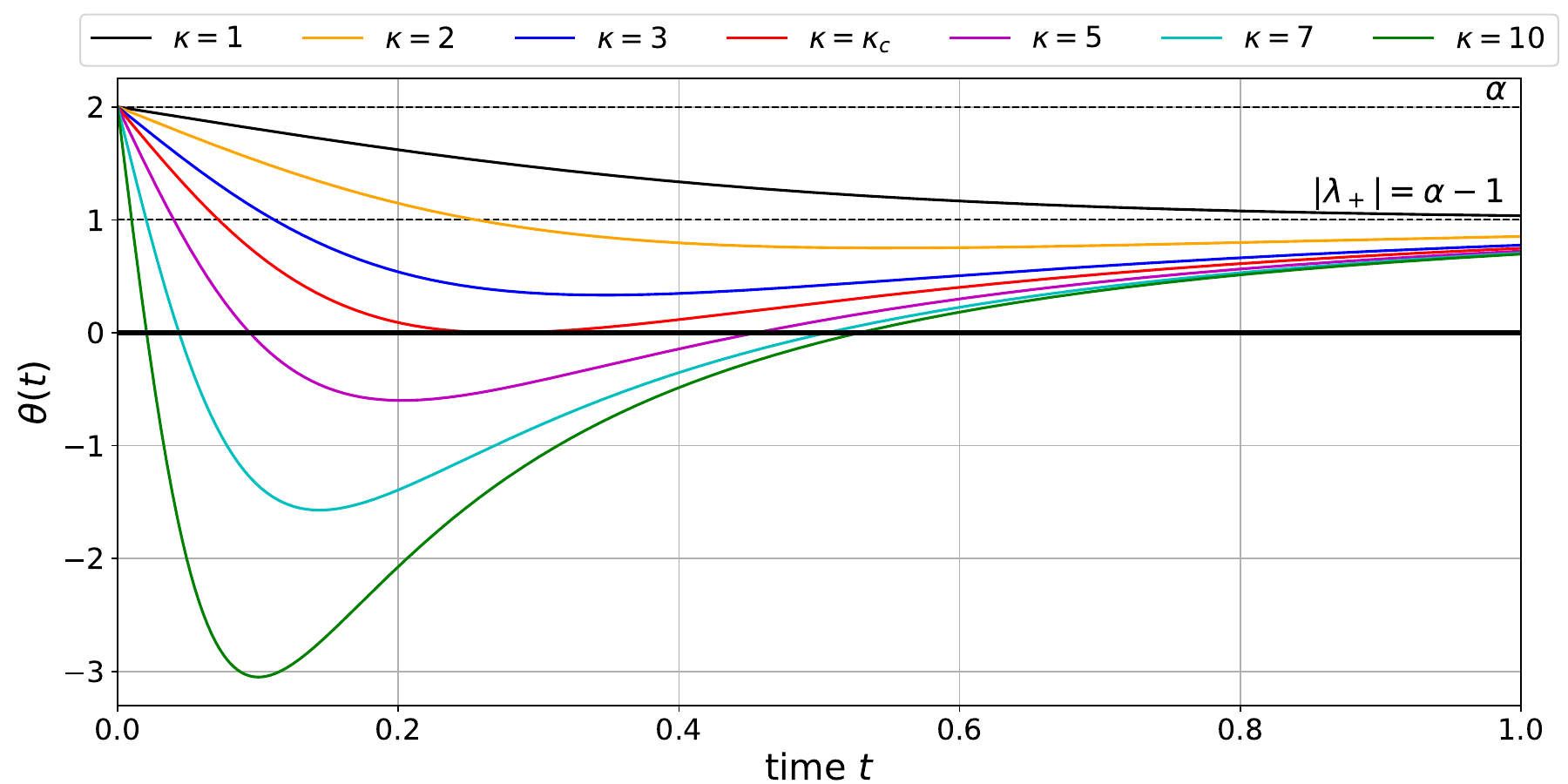}
    \caption{
        Mean-reversion rate $\theta_t$ of the reaction mode for different values of $\kappa$,
        for $\kappa$ going linearly from $1$ to $100$.
    }
    \label{fig:theta_t}
\end{figure}

Hence, the reduced non-normal system \eqref{eq:apx_nnf} locally experiences pseudo-critical intervals, characterised by $\Omega(t)<0$,
whenever $\kappa$ exceeds the above bounds.
This establishes that criterion \textbf{E1} (``approach to the unit root'') is reproduced purely by non-normality,
without any change in the spectrum, as illustrated in Figure \ref{fig:theta_t}.

\subsection{Proof: Diverging Variance}

The divergence of the variance in a non-normal system follows directly from the reduced reaction dynamics \eqref{eq:apx_reaction_rnnf}.
The time-dependent variance is
\begin{equation}
\bigl\langle r_t^{2}\bigr\rangle
~=~
2\delta~
\int_{0}^{t}
e^{-2\alpha s}
\bigl[\kappa^{2}\sinh(s)^{2}+\cosh(s)^{2}\bigr]~ds,
\end{equation}
so that the long–time (stationary) variance becomes
\begin{equation}
\mw{r^{2}}
~:=~
\lim_{t\to\infty}\mw{r_t^{2}}
~=~
\frac{\delta\alpha}{\alpha^{2}-1},
\frac{\kappa^{2}+1}{2}
\biggl[
1-
\frac{\kappa^{2}-1}{\kappa^{2}+1},
\frac{\alpha^{2}-1}{\alpha^{2}}
\biggr].
\end{equation}
The leading term grows as $\mathcal{O}(\kappa^{2})$,
demonstrating that $\mw{r^{2}}$ increases quadratically with the degree of non-normality.
Hence, a non-normal system naturally exhibits the early-warning signal \textit{E2} of increasing variance when $\kappa$ increases.

\subsection{Proof: Critical Slowing Down}

\begin{figure}
    \centering
    \includegraphics[width=0.75\textwidth]{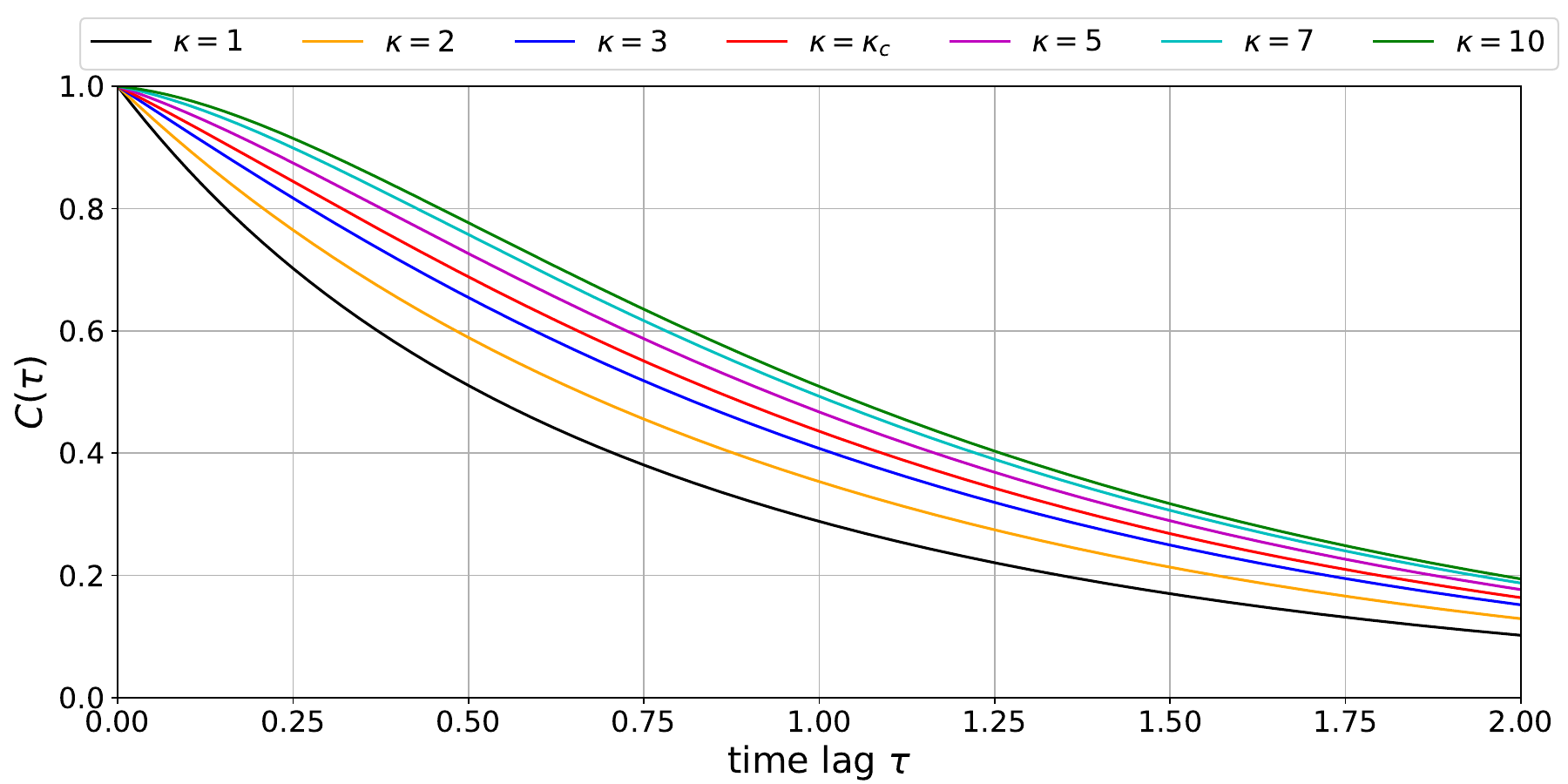}
    \caption{
        Lead-lag covariance $C_r(\tau)$ of the reaction mode $r_t$ for different levels of $\kappa$,
        for $\kappa$ going linearly from $1$ to $100$.}
    \label{fig:autocorr}
\end{figure}

As with the variance, proving that non-normal systems exhibit critical slowing down is straightforward.
Consider the reduced non-normal form along the reaction mode given by \eqref{eq:apx_reaction_red}, then
the covariance of the reaction component of the system is
\begin{equation}
    \mw{r_t r_{t+\tau}} ~=~
    2\delta ~ \int_0^t e^{-2\alpha t - \alpha\tau}
    \left[
    \kappa^2 \sinh(t+\tau)\sinh(t) + \cosh(t+\tau)\cosh(t)
    \right]~dt,
\end{equation}
from which the asymptotic covariance becomes
\begin{equation}
    \text{Cov}[r](\tau) = \lim_{t\to\infty} \mw{r_t r_{t+\tau}} =
    \frac{\delta\alpha}{\alpha^2 - 1}\frac{\kappa^2 + 1}{2}
    \left[
    1 - \frac{\kappa^2 - 1}{\kappa^2 + 1}
    \frac{\alpha^2 - 1}{\alpha^2}
    \right]
    e^{-\alpha\tau}
    \left[
    \cosh(\tau) + \frac{1}{\alpha}
    \frac{1}{1 - \frac{\kappa^2 - 1}{\kappa^2 + 1}
    \frac{\alpha^2 - 1}{\alpha^2}} \sinh(\tau)
    \right],
\end{equation}
which yields the asymptotic correlation function
\begin{equation}
C(\tau) = \frac{\text{Cov}[r](\tau)}{\mw{r^2}} =
e^{-\alpha\tau}
\left[
\cosh(\tau) + \frac{1}{\alpha}
\frac{1}{1 - \frac{\kappa^2 - 1}{\kappa^2 + 1}
\frac{\alpha^2 - 1}{\alpha^2}} \sinh(\tau)
\right].
\end{equation}
We can define the characteristic decay time $\tau_0$ as
\begin{subequations}
\begin{align}
\tau_0 &~~=~~
\left[-\left.\frac{d}{d\tau}\ln C(\tau)\right|_{\tau=0}\right]^{-1}
~~=~~ \left[
\frac{1}{\alpha}
\left(
\alpha^2 - \frac{1}{1 - \frac{\kappa^2 - 1}{\kappa^2 + 1}
\frac{\alpha^2 - 1}{\alpha^2}}
\right)
\right]^{-1} \\[1em]
\Rightarrow\qquad
\tau_0 &~~=~~ \frac{\mw{r^2}}{\delta} ~~=~~
\frac{\alpha}{\alpha^2 - 1}\frac{\kappa^2 + 1}{2}
\left[
1 - \frac{\kappa^2 - 1}{\kappa^2 + 1}
\frac{\alpha^2 - 1}{\alpha^2}
\right],
\end{align}
\end{subequations}
so that the correlation time diverges for $\kappa \to \infty$,
demonstrating that non-normal systems reproduce the early-warning signal \textit{E3} (critical slowing down),
as illustrated in Figure \ref{fig:autocorr}.

\subsection{Summary of Results}

In this section, we have shown that, along the reaction mode,
non-normal dynamics exhibit the same early-warning signals as a system approaching a bifurcation.
These findings can be summarised as follows:
\begin{itemize}
    \item[E1] \textbf{Drift toward a unit root}:
    The reaction dynamics can be written as an effective Ornstein-Uhlenbeck process with a time-dependent mean-reverting rate:
    \begin{equation}
        \dot{r}_t ~=~ -\theta(t)~r_t ~+~ \sqrt{2\delta}~\eta_t,
        \qquad \eta_t~\overset{\textit{i.i.d}}{\sim}~\mathcal{N}(0,1),
    \end{equation}
    such that for $\kappa \ge \kappa_c$ \eqref{eq:kappa_c_exact},
    there exists a time interval during which $\theta_t < 0$,
    meaning the system becomes locally critical.
    \item[E2] \textbf{Diverging variance}:
    The stationary variance of the reaction is given by $\mw{r^2} \sim \delta\kappa^2\alpha/(\alpha^2 - 1)$,
    which increases with $\kappa$ and diverges in the limit of large $\kappa$.
    \item[E3] \textbf{Critical slowing down}:
    The lag-$\tau$ autocorrelation function is $C(\tau) = e^{-\tau/\tau_c + \mathcal{O}((\alpha\tau)^2)}$,
    where the characteristic decay time $\tau_0 \sim \kappa^2$ increases with $\kappa$,
    indicating progressively slower recovery from perturbations.
\end{itemize}


\section{Degree of Non-Normality}


A degree of non-normality quantifies how far a matrix is from being normal. Specifically,
it increases as the eigenvectors become increasingly non-orthogonal, as the angle between them progressively narrows.
Consequently, the condition number $\kappa$ of the eigenbasis is a natural measure of non-normality.
However, other metrics also exist, such as the Henrici departure from normality and the Kreiss constant.

In this section, we first provide an intuitive and didactic example to illustrate how transient deviations emerge in a non-normal system.
In the second part, we demonstrate that the condition number $\kappa$ serves as an equivalent metric to the Henrici departure from normality.

\subsection{Intuition behind Transients in Two Dimensions}

Here we analyze system \eqref{eq:system} for the special case where $\delta=0$ in order to gain some intuition 
on the transient perturbations that push the system away from the equilibrium.

\begin{figure}[!htb]
    \centering
    \includegraphics[width=0.75\linewidth]{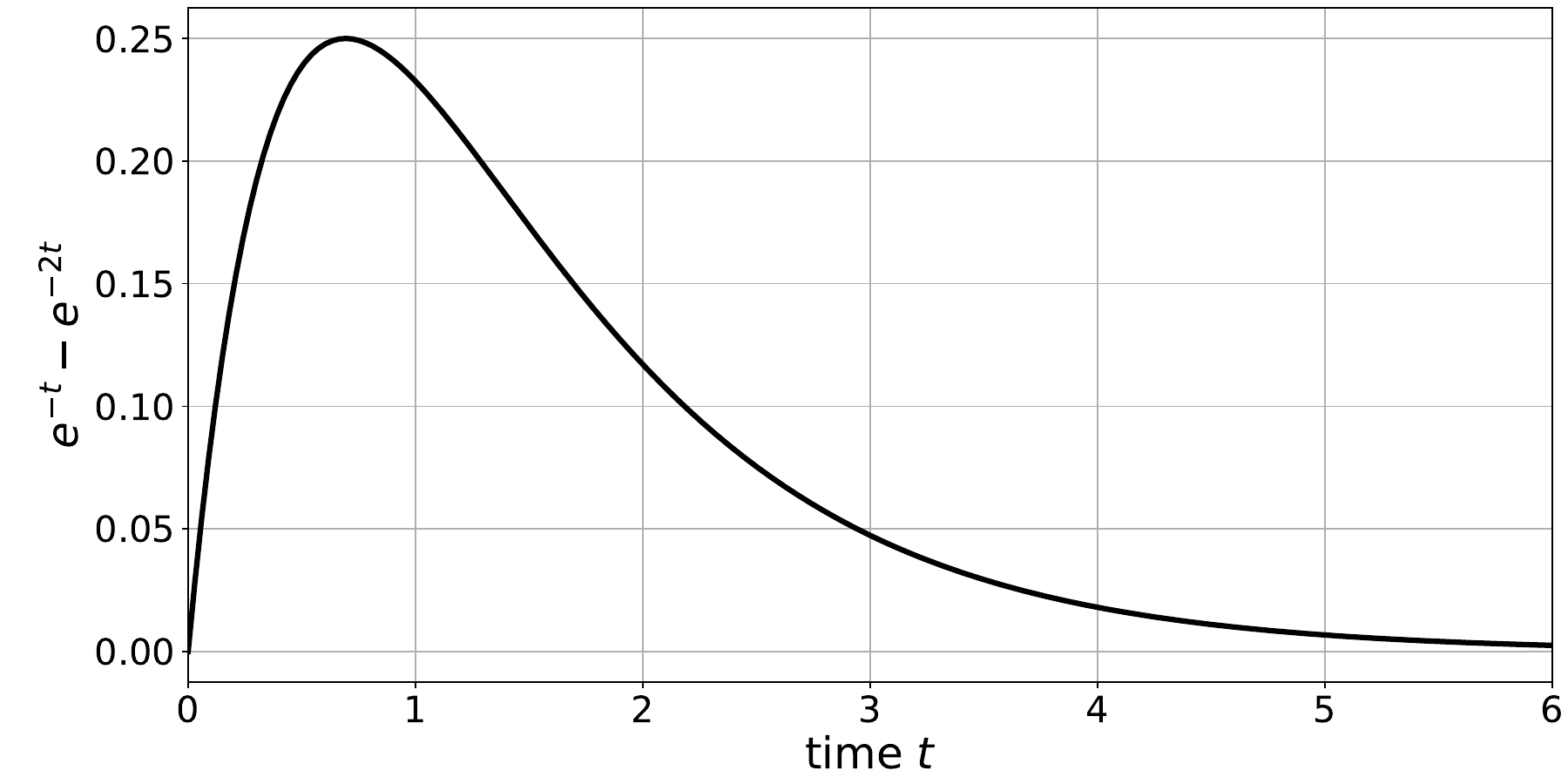}
    \caption{Function $e^{-t} - e^{-2t}$, characterising the essence of the non-monotonic behavior of a non-normal system relaxation to the equilibrium.}
    \label{fig:exp_diff_example}
\end{figure}

The deterministic solution of the reduced system \eqref{eq:apx_reaction_red} is given by
\begin{equation}
    \begin{pmatrix}
        n_t \\ r_t
    \end{pmatrix}
    =
    e^{\Gam t}
    \begin{pmatrix}
        n_0 \\ r_0
    \end{pmatrix},
    \quad\text{where }
    e^{\Gam t} = 
    e^{-\alpha t}
    \begin{pmatrix}
        \cosh(t) & \kappa^{-1}\sinh(t) \\
        \kappa\sinh(t) & \cosh(t)
    \end{pmatrix}
\end{equation}
where $(n_0\,,\, r_0)$ defines the initial position.
A transient emerges when the matrix norm $\|e^{\Gam t}\|$ increases in time.
Using the $L_2$-norm and applying the triangle inequality, the following 
upper bound to the exponential matrix norm is obtained:
\begin{equation}    \label{eq:SVD_exp}
    \|e^{\Gam t}\| \le e^{-\alpha t}\left[\cosh(t) + \kappa\sinh(t)\right]
    ,\quad t>0.
\end{equation}
When the system is normal, $\kappa = 1$,
and the the matrix norm is decreasing according to the largest eigenvalue $\lambda_{+} = -\alpha + 1$.
When the system is non-normal i.e. $\kappa>1$;
a new term appears in the upper bound which involves $e^{-t}$,
leading to a linear combination of two exponential $e^{-(\alpha+1)t}$ and $e^{-(\alpha-1)t}$.
When $\kappa\gg 1$, the leading order term is given by $\kappa(e^{\lambda_+t} - e^{\lambda_-t})$,
where $\lambda_\pm = -\alpha \pm 1$.
The function $e^{\lambda_+t} - e^{\lambda_-t}$ is non-monotonic and exhibits a maximum at $t^* = \ln(\lambda_+/\lambda_-)/(\lambda_+ - \lambda_-)$ as shown in Figure \ref{fig:exp_diff_example}. 
Therefore, the exponential matrix norm is not necessarily strictly decreasing as long as $\kappa$ is large enough.
In fact, it is this initial amplification that underlies the emergence of transient bursts in non-normal systems.

\subsection{Asymptotic Behavior of the Henrici Departure from Normality}

The degree of non-normality of a matrix $\A$ is typically measured by means of Henrici's departure from normality, defined as 
\begin{equation}
    d_F(\A)
    = \frac{\sqrt{\|\A\|_F^2 - \sum_i |\lambda_i|^2}}{\|\A\|_F}
\end{equation}
where $\| \cdot \|_F$ denotes the Frobenius norm.
We now re-write $d_F$ for the case where $\A$ is approximately normal along all directions (that is, all singular values are approximately one) with the exception of the last component
(see previous section for details about this assumption). 
Above, we have expressed the system's non-normality conveniently via its condition number $\kappa$. 
Here we further show that the behavior of $d_F$ is also dominated by $\kappa$. 

Assuming that the eigenbasis is nearly unitary in all directions except the last one, the transformation matrix \(\PP\) can be expressed as
\begin{equation}
    \PP = \mathbf{U}\Sig\V^\dag
    \quad\text{with}\quad
    \Sig = \mathbf{I} - (1 - \kappa^{-1})\mathbf{E}_{NN}
\end{equation}
and \(\mathbf{E}_{NN}\) is an $N \times N$ matrix that has all its elements equal to zero except for its $(N,N)$-th component which is equal to one. 

We now expand the eigenbasis \(\PP\) and its inverse in terms of \(\kappa\)
\begin{equation}
    \PP = \mathbf{Q}_0 - (1 - \kappa^{-1})\mathbf{Q}_N
    \quad\text{and}\quad
    \PP^{-1} = \mathbf{Q}_0^\dag - (1 - \kappa)\mathbf{Q}_N^\dag
\end{equation}
where \(\mathbf{Q}_0 = \U\V^\dag\) is a unitary matrix,
and \(\mathbf{Q}_N = \hu_N\hv_n^\dag\).
Therefore, $\A$ becomes 
\begin{equation}
    \A
    = \PP\Lam\PP^{-1}
    = \mathbf{Q}_0\Lam\mathbf{Q}_0^\dag
    - (1 - \kappa^{-1})\mathbf{Q}_N\Lam\mathbf{Q}_0^\dag
    - (1 - \kappa)\mathbf{Q}_0\Lam\mathbf{Q}_N^\dag
    - (1 - \kappa^{-1})(1 - \kappa)\mathbf{Q}_N\Lam\mathbf{Q}_N^\dag.
\end{equation}
Since the Frobenius norm is invariant under unitary transformations,
we compute it in the basis \(\mathbf{Q}_0\), 
\begin{equation}
    \A_Q
    = \mathbf{Q}_0^\dag \A\mathbf{Q}_0
    = \Lam - (1 - \kappa^{-1}) \V_N \Lam
    - (1 - \kappa) \Lam \V_N
    - (1 - \kappa^{-1})(1 - \kappa) \V_N \Lam \V_N,
    \label{thtgfq1}
\end{equation}
where \(\V_N = \mathbf{Q}_0^\dag \mathbf{Q}_N = \mathbf{\hat{v}}_N \mathbf{\hat{v}}_N^\dag\) is a projection matrix and we have used that
\(\V_N = \V_N^\dag\) and \(\V_N^2 = \V_N\).
The conjugate transpose \(\A_Q^\dag\) reads 
\begin{equation}
    \A_Q^\dag
    = \Lam^\dag - (1 - \kappa^{-1}) \Lam^\dag \V_N
    - (1 - \kappa) \V_n \Lam^\dag
    - (1 - \kappa^{-1})(1 - \kappa) \V_N \Lam^\dag \V_N ~.
    \label{thtgfq}
\end{equation}

The Frobenius norm \(\|\A_Q\|_F^2\) is given by
\begin{equation}
    \|\A_Q\|_F^2 = \text{Tr}\left[\A_Q^\dag \A_Q \right]. 
\end{equation}
Using expressions (\ref{thtgfq1}) and (\ref{thtgfq}), 
the Frobenius norm of \(\A\) is equal to
\begin{equation} \label{eq:A_F_intermediate}
    \|\A\|_F^2
    = \|\A_Q\|_F^2
    = \|\Lam\|_F^2
    + \text{Tr}\left[\Lam\Lam^\dag \V_N\right]f_1(\kappa)
    + \text{Tr}\left[\Lam\V_N \Lam^\dag \V_N \right]f_2(\kappa)
\end{equation}
where we used the cyclic property of the trace and defined
\begin{equation}
   f_1(\kappa) = (\kappa - \kappa^{-1})^2
   \quad\text{and}\quad
   f_2(\kappa) = 3\kappa^2 (1 - \kappa^{-1})^4. 
\end{equation}
Expression \eqref{eq:A_F_intermediate} can further be simplified by noting that 
\begin{subequations}
\begin{align}
    \text{Tr}\left[\Lam \Lam^\dag \V_N \right]
    &= \sum_{i=1}^N |\lambda_i|^2 |v_{Ni}|^2 
    = \mw{|\lambda|^2}_w, \\
    \text{Tr}\left[\Lam\V_N \Lam^\dag \V_N \right]
    &=  \left|\sum_{i=1}^N \lambda_i |v_{Ni}|^2 \right|^2
    = \left|\mw{\lambda}_w \right|^2.
\end{align}
\end{subequations}
Additionally, the Frobenius norm of \(\Lam\) is simply given by 
\begin{equation}
    \|\Lam\|_F^2
    = \sum_{i=1}^N |\lambda_i|^2
    = N\mw{|\lambda|^2}.
\end{equation}
Putting all the above together, we obtain the following expression for Henrici's departure from normality: 
\begin{equation}
    d_F(\A) 
    = 
    \sqrt{\frac{\mw{|\lambda|^2}_w f_1(\kappa) + \left|\mw{\lambda}_w \right|^2 f_2(\kappa)}{N\mw{|\lambda|^2} + \mw{|\lambda|^2}_w f_1(\kappa) + \left|\mw{\lambda}_w \right|^2 f_2(\kappa)}}. 
\end{equation}
For $\kappa$ close to $1$, that is for the system close to normal, we define $\epsilon = \kappa -1 \gtrsim 0$ and obtain to first order
\begin{equation}
    d_F(\A) = 2\sqrt{\frac{\mw{|\lambda|^2}_w}{N \mw{|\lambda|^2}}} ~\epsilon + \mathcal{O}\left(\epsilon^3\right), 
\end{equation}
showing that Henrici's departure from normality decays linearly to zero as a function of $\kappa -1$ as anticipated. 
By contrast, when $\A$ is strongly non-normal, that is $\kappa \gg 1$, the Henrici departure from normality becomes
\begin{equation}
    d_F(\A)
    = 1 - \kappa^{-2} \frac{N \mw{|\lambda|^2}}{\mw{|\lambda|^2}_w
    + 3 |\mw{\lambda}_w|^2}
    + \mathcal{O}(\kappa^{-3})
\end{equation}
and hence approaches the maximum non-normality $d_F=1$ as $1 - {\cal O}(\kappa^{-2})$. 

\begin{figure}
    \centering
    \includegraphics[width=0.5\textwidth]{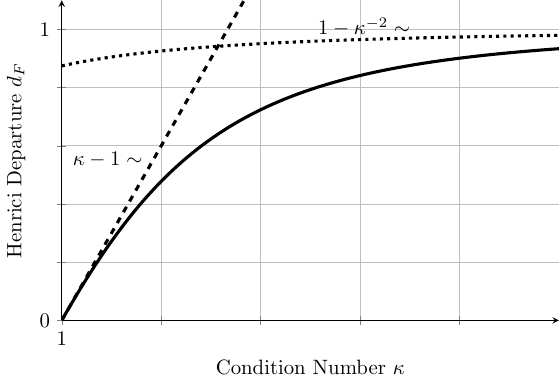}
    \caption{
        Henrici departure from normality as a function of the condition number \(\kappa\),
        where the dashed straight line represent the asymptotic behavior close to normality ($0 <\kappa-1 \ll  1$);
        the dotted line gives the asymptotic behavior when the system tends to be highly non-normal
        $\kappa\gg 1$.
    }
    \label{fig:henrici_departure}
\end{figure}

\section{Empirical Study \& Simulation}
\label{SI:empirical}

Here we describe how to measure the non-normal component of a matrix $\A$ governing system \eqref{eq:system}.

\subsection{Measuring Non-Normal and Reaction Mode From Observations}

Let us consider the discretized version of \eqref{eq:system} given by 
\begin{equation} \label{eq:system_discrete}
    \mathbf{x}_{t+1}
    = \mathbf{A}'\mathbf{x}_t + \sqrt{2\delta \Delta t} ~\boldsymbol{\eta}_t
    \quad \text{where }
    \boldsymbol{\eta}_t \overset{i.i.d.}{\sim} \mathcal{N}(0,\mathbf{I})
    \quad\text{and }
    \A' = \mathbf{I} + \Delta t\A
\end{equation}
and we assume that its states $\x_1, \x_2, \ldots, \x_T$ are observed at equidistant times. 
To estimate the matrix \(\mathbf{A}'\), 
we stack these observations into matrices \(\mathbf{X} = (\mathbf{x}_1,\ldots,\mathbf{x}_{T})\) and \(\mathbf{Y} = (\mathbf{x}_2,\ldots,\mathbf{x}_T)\).
The interaction matrix \(\mathbf{A}'\) can be extracted by solving the least squares problem \(\mathbf{Y} = \mathbf{A}'\mathbf{X}\) with solution \(\widehat{\mathbf{A}'} = \mathbf{Y}\mathbf{X}^{+}\), where \(\mathbf{X}^{+}\) is the pseudo-inverse of \(\mathbf{X}\).
In the limit where $T$ is very large, this allows for a faithful reconstruction of the true matrix $\A$. 
However, in practice, especially when the ratio of $N/T$ is comparatively high, the estimate $\hat{\A}$ is error prone.

In principle, once $\A$ is estimated, one can then estimate the non-normal component $\hat{\mathbf{n}}$ and reaction component $\hat{\mathbf{r}}$ via SVD of $\hat{\A}$ (see Section \ref{SI:dimension_reduction}).
However, estimation errors of $\A$ will further be amplified by the SVD, leading to imprecise estimations of the non-normal and reaction components. 
These errors are exacerbated when the condition number $\kappa$ of the matrix is high, making the matrix ill-conditioned and sensitive to small perturbations in the input. 

Therefore, we propose to estimate the non-normal and reaction component via an optimization-based approach that is more robust to numerical errors.
Noting that \(\mathbf{A}' \sim  \kappa \mathbf{\hat{r}} \mathbf{\hat{n}}^\dag + \mathcal{O}(1)\)
in highly non-normal systems,
we reformulate the problem as
\begin{equation} \label{eq:quadratic_optimization}
    \hat{\mathbf{r}}, \hat{\mathbf{n}}
    =
    \argmax_{ \mathbf{v}, \mathbf{w} \in \mathbb{S}^{n-1} } \langle \mathbf{v}, \hat{\A}' \mathbf{w} \rangle  
    ~~ \text{s.t.}~~
    \langle \mathbf{v}, \mathbf{w} \rangle = 0
\end{equation}
where $\mathbb{S}^{n-1}$ is the unit sphere in $n$ dimensions.
Problem \eqref{eq:quadratic_optimization} is a standard quadratic convex optimization problem and readily solvable with standard gradient descent methods.

\begin{figure}[ht]
    \centering
    \begin{subfigure}{0.48\textwidth}
        \centering
        \includegraphics[width=\textwidth]{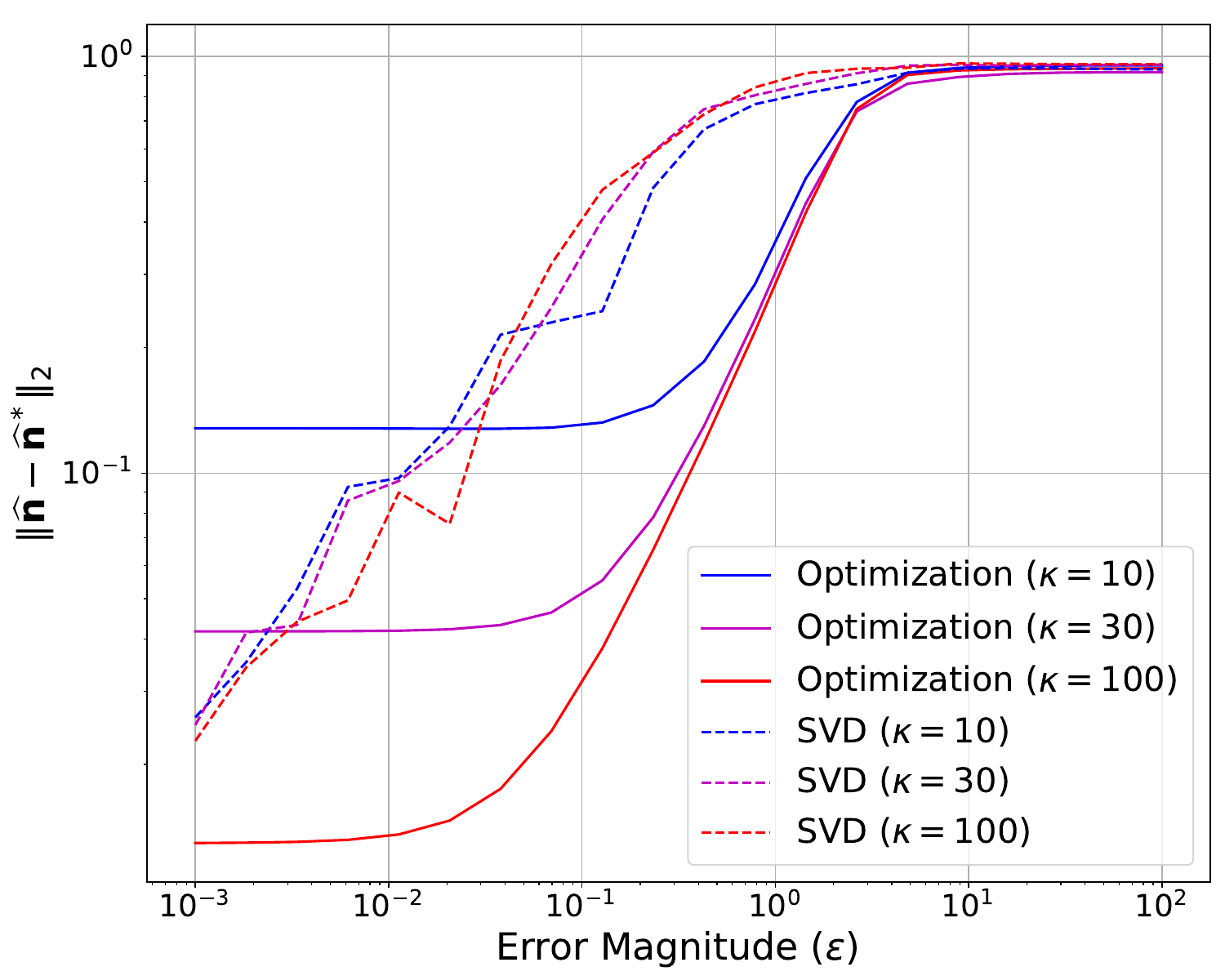}
    \end{subfigure}
    \begin{subfigure}{0.48\textwidth}
        \centering
        \includegraphics[width=\textwidth]{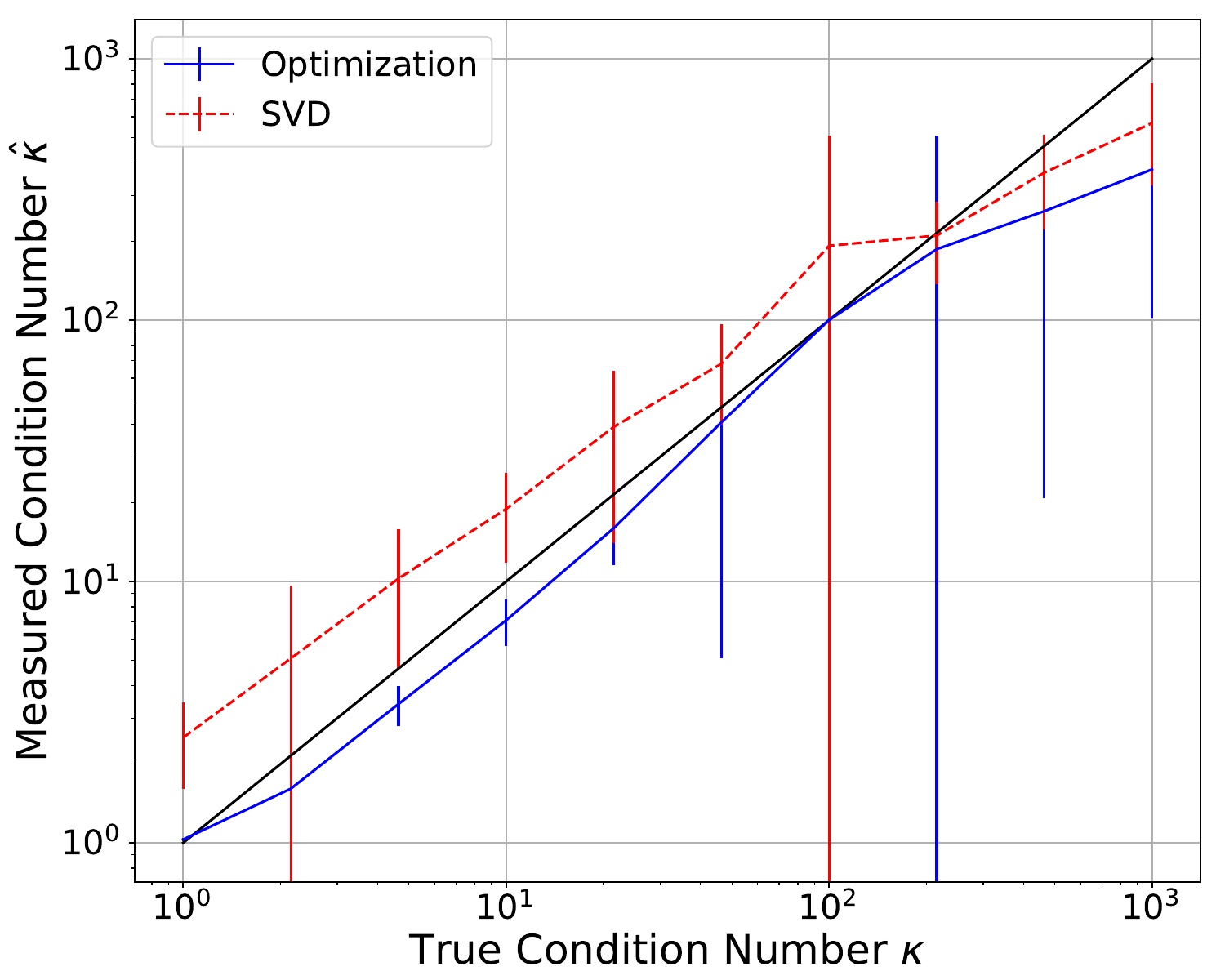}
    \end{subfigure}
    \caption{
       	(Left)
        L2 norm between true non-normal mode $\hat{\mathbf{n}}$ and estimated non-normal mode $\hat{\mathbf{n}}^*$ for different levels of noise $\epsilon=\|\mathbf{E}\|_2/\|\mathbf{A}\|_2$.
        The solid lines represent errors from estimates via the optimization procedure \eqref{eq:quadratic_optimization}
        while dashed lines indicate errors from estimates via SVD. 
        The maximum error is one, which coincides with an estimate that is fully orthogonal to the ground truth one.
        (Right)
        Measured condition number $\hat{\kappa}$ as a function of the true condition number $\kappa := \kappa(\PP)$ of the matrix $\PP$ diagonalizing the matrix $\A$. 
        We compare our method \eqref{eq:meas_kappa}, which bypasses the need to directly measure $\kappa$ via singular values extracted from the SVD, to the method using the SVD.
    }
    \label{fig:svd_vs_opt_noise}
\end{figure}

We now compare the performance of optimization \eqref{eq:quadratic_optimization} with the direct calculation of $\hat{\mathbf{n}}$ via SVD.
To this end, we generate noisy matrices $\A + \mathbf{E}$ where $\A$ is the known ground-truth and $\mathbf{E}$ is a Gaussian matrix. 
We scale the matrix such that $\epsilon = \|\mathbf{E}\|/\|\A\|$ is the amplitude of the perturbation, allowing us to control the relative level of noise.
We then measure $\hat{\mathbf{n}}$ from $\A + \mathbf{E}$ both via \eqref{eq:quadratic_optimization} and via SVD, and measure the difference between true non-normal mode $\hat{\mathbf{n}}$ and estimated non-normal mode $\hat{\mathbf{n}}^*$ with the $L2$-norm
$\left| \left| \hat{\mathbf{n}} - \hat{\mathbf{n}}^* \right| \right|$. 
Plotting this error as a function of the noise scale $\epsilon$ in Figure \ref{fig:svd_vs_opt_noise} (left), 
we notice that, especially for large noise and strongly non-normal systems $\kappa$, our optimization estimates are more robust, demonstrating resilience to noise levels up to an order of magnitude greater.

In principle, once the matrix $\hat{\A}$ has been measured by calibrating a VAR(1) process, 
one could proceed to estimate its non-normality by first determining the transition matrix $\PP$ that diagonalizes $\hat{\A}$,
and subsequently calculating the SVD of $\PP=\mathbf{U} \Sigma \mathbf{V}^\dagger$ to identify the non-normal mode $\hat{\mathbf{n}} \sim \mathbf{u}_n$ and reaction mode $\hat{\mathbf{r}} \sim \sum_{i=1}^{n-1} \mathbf{u}_i$.
However, as detailed above, this approach is highly sensitive to noise, especially when $\A$ is non-normal.
Therefore, we note that, from \eqref{eq:apx_nnf}, the condition number $\kappa$ can be estimated directly from $\hat{\mathbf{r}}$ and $\hat{\mathbf{n}}$ via
\begin{equation}    \label{eq:meas_kappa}
    \kappa = \sqrt{\left| \frac{\langle \hat{\mathbf{r}}, \hat{\A} \hat{\mathbf{n}} \rangle}{\langle\hat{\mathbf{n}}, \hat{\A} \hat{\mathbf{r}} \rangle} \right|}. 
\end{equation}

In Figure \ref{fig:svd_vs_opt_noise} (right), we compare the difference in the estimated $\kappa$ when using \eqref{eq:meas_kappa} versus directly calculating it from the SVD, 
and find that our approach is generally more accurate up to $\kappa \approx 200$. 
Both methods tend to underestimate $\kappa$ as the system's non-normality becomes extremely large.

\subsection{Measuring Quasi-Deterministic Loops}

We identify quasi-deterministic cycles and their amplitude as follows.

First, we project the discretized dynamics $\mathbf{x}_1, \mathbf{x}_2, \ldots \mathbf{x}_T$ onto the normal-mode $\hat{\mathbf{n}}$ and reaction-mode $\hat{\mathbf{r}}$, giving rise to a two-dimensional trajectory $(n_1, r_1), (n_2, r_2), \ldots, (n_T, r_T)$. 
For a given starting point $(n_t, r_t)$ at time $t$, we then ask what is the first time for which the trajectory returns back to this point. 
To this end, we draw two circles $C_I$ and $C_O$ around $(n_t,r_t)$, one of radius $5B$ (inner circle $C_I$) and one of radius $10B$ (outer circle $C_O$) where $B$ is the (empirically measured) standard deviation of the fluctuations (providing the unit scale for these circles).
We define the departure from $(n_t, r_t)$ as the first time $\tau_1 > t$ at which the trajectory leaves $C_O$, and the return time as the first time $\tau_2 > \tau_1$ at which the trajectory returns back into circle $C_I$.
The trajectory $\{ (n_t,r_t) \}_{t=\tau_1}^{\tau_2}$ then defines one cycle, and we measure its amplitude as its diameter.

We then measure all cycles in terms of the above cycle trajectories by iterating over all possible starting points along a given trajectory.

\subsection{Dynamical Interpretation of Quasi-Deterministic Loops}

From system \eqref{eq:apx_nnf} we deduce that 
\(dr_t \approx (\kappa n_t - \alpha_r r_t) dt\)
and distinguish between the expansion phase, $dr_t /r_t > 0$, during which the transient grows, and contraction phase $d r_t / r_t < 0$ when the system relaxes back to equilibrium. 
The non-normal mode, meanwhile, is governed by the mean-reversion process
\(dn_t \approx -\alpha_n n_t dt + \sigma dW^n_t\)
where we do not ignore the stochastic component because it is not necessarily negligible. 
Taken together, this yields 
\begin{equation}
    r_t
    \approx \kappa\int^{t}_0 e^{-\alpha_r(t-s)}n_s ds
    = \kappa\sigma\int^{t}_0 e^{-\alpha_r(t-s)}\int^{s}_0e^{-\alpha_n(s-u)} dW^n _u \, ds
\end{equation}
and suggests that the Quasi-Deterministic loops arise from the exponential smoothing of the non-normal mode along the reaction mode,
with the reaction dynamics driven by a mean-reverting OU process.

\section{Relaxing the Assumptions Regarding Singular Values}
\label{SI:general_SV}

Throughout the manuscript, we have assumed that $\sigma_1 \approx \sigma_2 \approx \ldots \approx \sigma_{N-1} \approx 1 \gg \sigma_{N} = \kappa^{-1}$. 
The low-rank hypothesis in complex networks supports this assumption,  stating that many high-dimensional nonlinear systems can be reduced to lower-dimensional linear systems.
In numerous sociotechnical network systems, only a ``few'' singular values from a Singular Value Decomposition (SVD) of the network are large,
while most middle singular values are ``almost'' the same, and a final ``few'' are much smaller.
In our framework, these smaller singular values correspond to multiple non-normal modes driving transient deviations.

To further clarify this point,  
we consider the generic case where the singular values are independently Gaussian distributed,  
i.e., $\sigma_i \overset{\text{i.i.d.}}{\sim} |\mathcal{N}(0, 1)|$ where the absolute value ensures positivity. 
It is well known that the ratio between any two such singular values follows a Cauchy distribution with the probability density function (PDF)  
\begin{equation}
p(x) \sim \frac{1}{x^{1+\alpha}}, \quad \text{with } \alpha = 1.
\label{ghwtrbgqg}
\end{equation}
This distribution implies undefined variance and mean,  
mathematically, and captures the divergence where the largest singular values dominate significantly.  

For instance,  without loss of generality, let us order the singular values as  
\[
\sigma_0 \geq \cdots \geq \sigma_n \geq \cdots \geq \sigma_{N} > 0,
\]
then the ratio $\sigma_0 / \sigma_n$ grows linearly with $n$.

\begin{figure}
    \centering
        \begin{subfigure}{0.495\textwidth}
        \centering
        \includegraphics[width=\textwidth]{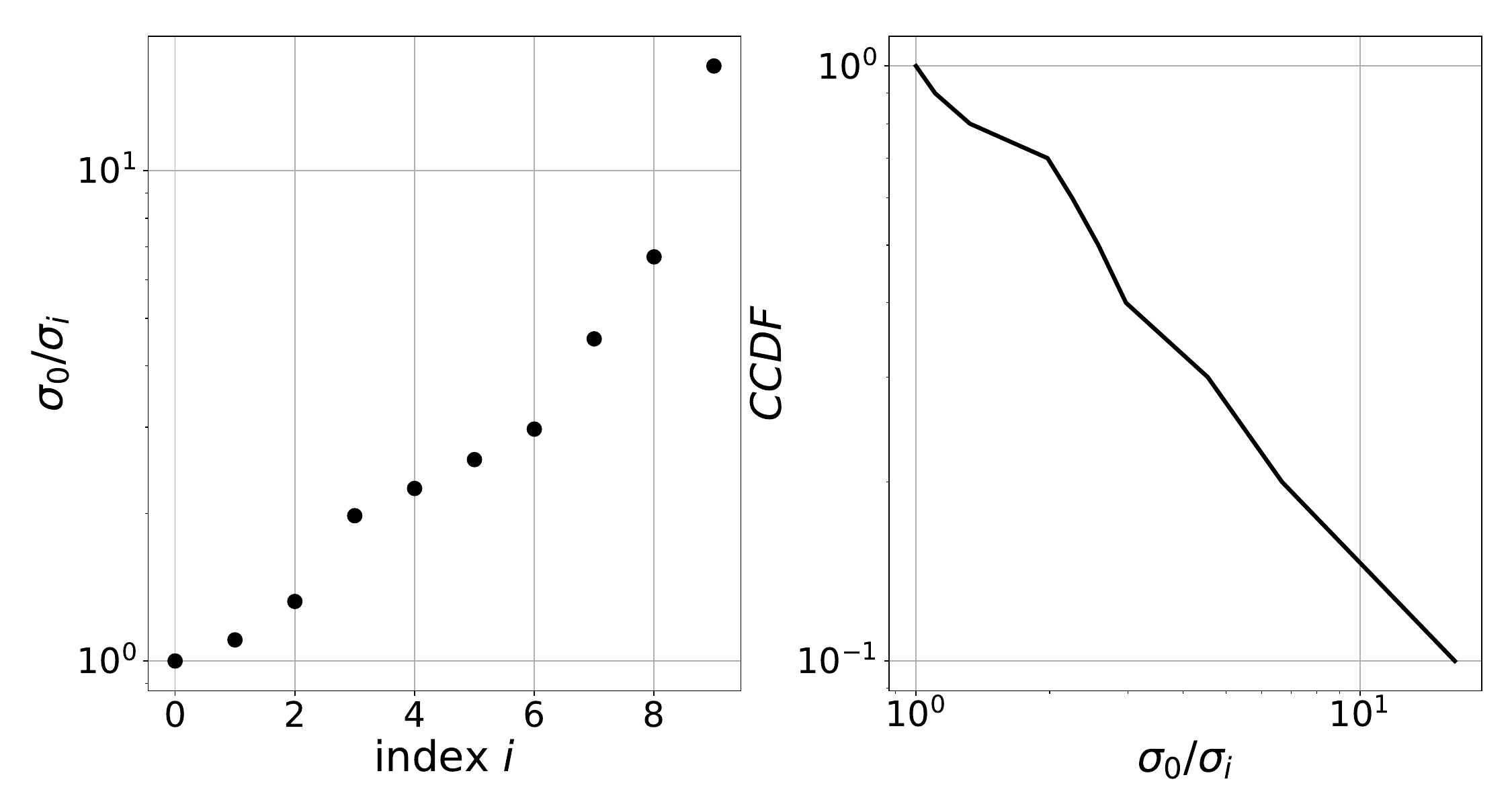}
        \end{subfigure}    
        \begin{subfigure}{0.495\textwidth}
        \centering
        \includegraphics[width=\textwidth]{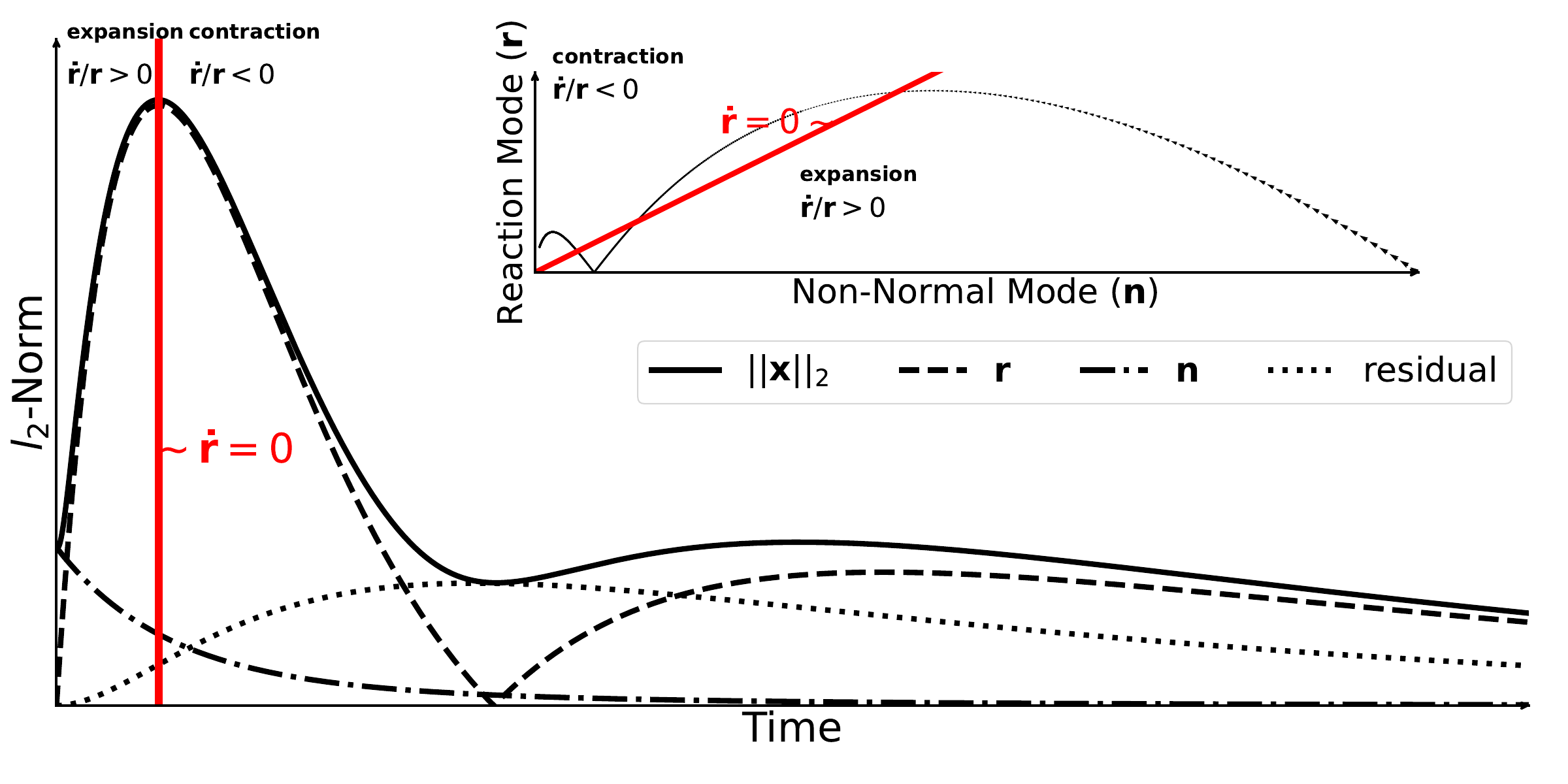}
        \end{subfigure}
        \caption{
        The plot on the right is the same as Figure 1 in the main manuscript, 
        but generated for an $N=10$-dimensional system in which all singular values are sampled from $|\mathcal{N}(0, 1)|$. 
        As evidenced in the left and center plot, the singular value ratios are Cauchy-distributed, and the non-normality is chiefly driven by 
        $\sigma_9 / \sigma_{10}$, which plays a dominant role in shaping the transient response. as seen in the right most plot. 
        }
    \label{fig:gaussian_svd}
\end{figure}

To illustrate that this generic setup gives rise to non-normal transients described by our formalism,
we sample singular values according to the above procedure for $N=10$, and 
assume the same eigenvalues as in the main manuscript ($\lambda_i = -i$). 
The left-most and center plot of Figure \ref{fig:gaussian_svd} confirm that the ratios of these singular values are indeed power law 
distributed according to (\ref{ghwtrbgqg}). 
The right-most plot shows the same graph as Figure 1 in our main article but with this adjusted set of singular values. 
Evidently, even with anisotropy in the singular values,  
highly non-normal and hierarchical networks do exhibit transients that arise within a reduced subspace of the system.  

Therefore, while our study adopts a simplified scenario in which the first $N-1$
singular values are approximately equal and the last is markedly smaller, the key result, transient dimensionality reduction, 
holds broadly across diverse non-normal linear systems.

\end{document}